\documentclass[submitting]{nst}

\usepackage{subfigure,dcolumn}
\usepackage[T2A,T1]{fontenc}
\usepackage[russian,english]{babel}

\usepackage{listings}
\usepackage{makecell}
\begin{document}

\title{Discrimination of \textit{pp} solar neutrinos and $^{14}$C double pile-up events in a large-scale LS detector}\thanks{This work was supported by National Natural Science Foundation of China No. 12005044, the Strategic Priority Research Program of the Chinese Academy of Sciences, Grant No. XDA10011200, Guangxi Science and Technology Program No. GuiKeAD21220037.}

\author{Guo-Ming Chen}
\affiliation{School of Physical Science and Technology, Guangxi University, Nanning 530004, China.}
\author{Xin Zhang}
\affiliation{Institute of High Energy Physics, Beijing 100049, China.}
\affiliation{University of Chinese Academy of Sciences, Beijing 100049, China.}
\author{Ze-Yuan Yu}
\affiliation{Institute of High Energy Physics, Beijing 100049, China.}
\author{Si-Yuan Zhang}
\affiliation{School of Physical Science and Technology, Guangxi University, Nanning 530004, China.}
\author{Yu Xu}
\affiliation{School of Physics, Sun Yat-Sen University, Guangzhou 510275, China.}
\author{Wen-Jie Wu}
\affiliation{Department of Physics and Astronomy, University of California, Irvine, California, USA}
\author{Yao-Guang Wang}
\affiliation{Institute of High Energy Physics, Beijing 100049, China.}
\author{Yong-Bo Huang}
\email[Yong-Bo Huang, ]{huangyb@gxu.edu.cn}
\affiliation{School of Physical Science and Technology, Guangxi University, Nanning 530004, China.}

\begin{abstract}
    As a unique probe, precision measurement of \textit{pp} solar neutrinos is important for studying the Sun's energy mechanism, monitoring thermodynamic equilibrium, and studying neutrino oscillation in the vacuum-dominated region. For a large-scale liquid scintillator detector, one bottleneck for \textit{pp} solar neutrino detection comes from pile-up events of intrinsic $^{14}$C decays. This paper presents a few approaches to discriminate \textit{pp} solar neutrinos and $^{14}$C pile-up events by considering the difference in their time and spatial distributions. In this work, a Geant4-based Monte Carlo simulation is constructed. Then multivariate analysis and deep learning technology were adopted respectively to investigate the capability of $^{14}$C pile-up reduction. As a result, the BDTG model and VGG network showed good performance in discriminating \textit{pp} solar neutrinos and $^{14}$C double pile-up events. Their signal significance can achieve 10.3 and 15.6 using only one day of statistics. In this case, the signal efficiency is 51.1\% for discrimination using the BDTG model when rejecting 99.18\% $^{14}$C double pile-up events, and the signal efficiency is 42.7\% for the case using the VGG network when rejecting 99.81\% $^{14}$C double pile-up events.
 
\end{abstract}


\maketitle

\section{Introduction}
\label{section:Introduction}
    
    With the development of nuclear physics and astrophysics, we were able to glimpse the Sun's energy mechanism, which comes from the nuclear fusion of light nuclei in the core of the Sun~\cite{Bethe-1938,Bethe-1939,Bahcall-1996pt}. The proton-proton (\textit{pp}) cycle produces $\sim$99\% of the solar energy, and its primary reaction is the fusion of two protons into a deuteron:

    \begin{large}
		\begin{equation} 
		p + p = {^{2}H} + e^{+} + \nu_{e}
		\label{eq:pp}
		\end{equation} 
	\end{large}
    
    In the reaction, large amounts of low-energy neutrinos are emitted, named \textit{pp} neutrinos ($E \textless 0.42$~MeV). In addition, the proton-electron-proton (\textit{pep}) process and the secondary reactions in the \textit{pp} cycle emit neutrinos as well, they are known as \textit{pep} neutrinos, $^{7}$Be neutrinos, $^{8}$B neutrinos and \textit{hep} (helium-proton) neutrinos, respectively. The remaining energy of the Sun is contributed by the carbon-nitrogen-oxygen (CNO) cycle, with CNO neutrinos emitted. The detection of solar neutrinos is considered as a direct way to test theoretical solar models. However, differences between early observations and theoretical predictions were discovered~\cite{Davis-1968cp, Cleveland-1998nv, GALLEX-1992gcp, GALLEX-1998kcz, Kaether-2010ag, GNO-2005bds, SAGE-2009eeu, Gavrin-2019sok, Kamiokande-II-1989hkh, Kamiokande-1996qmi}, leading to the so-called "solar neutrino problem" that has plagued us for more than 30 years. Later, the MSW-LMA mechanism~\cite{Wolfenstein-1977ue, Mikheyev-1985zog} was proved to be the standard solution since solid evidence was provided by SNO~\cite{SNO-2001kpb, SNO-2003bmh} and KamLAND~\cite{KamLAND-2002uet}. Currently, the Standard Solar Model (SSM)~\cite{Bahcall-1995bt, Christensen-Dalsgaard-1996hpz, DeglInnocenti-1996uex, Brun-1999dw, Bahcall-2001pf, Serenelli-2009yc} can provide a precise prediction for the flux and energy distribution of solar neutrinos. As for the detection of solar neutrinos, almost all solar neutrino components have been observed~\cite{Borexino-2007kvk, BOREXINO-2014pcl, BOREXINO-2018ohr, BOREXINO-2020aww}, and we are expected to enter an era of precise and comprehensive measurement of solar neutrinos in the next decades~\cite{Gann-2021ndb, Xu-2022wcq}.
   
    \textit{pp} neutrinos are strongly related to the predominant energy production of the Sun and carry the recent message of the core of the Sun. These characteristics make \textit{pp} neutrinos a unique messenger for the study of the Sun's energy mechanism and thermodynamic equilibrium monitoring. On the other hand, \textit{pp} neutrinos can be used for the study of neutrino oscillation in the vacuum-dominated region. The detection of \textit{pp} neutrinos requires a low threshold ($\sim 200$~keV) and effective background reduction at the same time. The first detection of \textit{pp} neutrinos was made by $^{71}$Ga-based radiochemical detectors~\cite{GALLEX-1992gcp, GALLEX-1998kcz, Kaether-2010ag, GNO-2005bds, SAGE-2009eeu, Gavrin-2019sok}. Later, a large-scale liquid scintillator (LS) detector was successfully applied in the Borexino experiment and provided the best measurement of \textit{pp} neutrinos at $\sim$10\% level~\cite{BOREXINO-2014pcl, BOREXINO-2018ohr} via elastic neutrino-electron scattering. 
    
    According to the experience from Borexino, intrinsic $^{14}$C decays from the organic liquid scintillator and its associated pile-up events are a crucial internal background for a large-scale LS detector. $^{14}$C pile-up events correspond to the case that more than one $^{14}$C decays at different detector positions but takes place in the same trigger window. In addition, pile-up can be classified into the following categories according to the multiplicity of $^{14}$C accidental coincidence: double pile-up, threefold pile-up, fourfold pile-up, and so on. The Borexino experiment ($\sim$278~ton) takes a lot of effort in LS purification and makes the $^{14}$C concentration reach about $2.7 \times 10^{-18}$~g/g. With this concentration, $^{14}$C double pile-up is about 10\% of the events in the spectral gap between $^{14}$C and $^{210}$Po spectra~\cite{BOREXINO-2014pcl}. 
   
    \begin{table*}[!htb]
	\caption{The event rates (unit:~cpd/kton) of \textit{pp} neutrinos, $^{14}$C single and pile-up events in different $^{14}$C concentrations. In this table, a spherical LS detector ($\sim$12~kton) with a 15~m radius is used in the calculation, and the time window is 500~ns. The values in the brackets indicate the event rates inside the energy range of interest (0.16, 0.25)~MeV, the ratio is about 10\% for both \textit{pp} neutrinos and $^{14}$C double pile-up events.}
	\label{table:tab-eventRate}
		\renewcommand\arraystretch{2.4}
		\setlength{\tabcolsep}{7mm}
		\centering
		\begin{tabular}{c|c|c|c|c}
			\hline
			 Event types & $10^{-18} g/g$ & \makecell[c]{$2.7\times10^{-18} g/g$ \\ (Borexino-like)} & $5\times10^{-18} g/g$ & $10^{-17} g/g$   \\ \hline
			\textit{pp}-$\nu$ & \makecell[c]{$1.37\times10^{3}$ \\ ($\sim$~$1.37\times10^{2}$~)} & \makecell[c]{$1.37\times10^{3}$  \\ ($\sim$~$1.37\times10^{2}$~)} & \makecell[c]{$1.37\times10^{3}$ \\ ($\sim$~$1.37\times10^{2}$~)} & \makecell[c]{$1.37\times10^{3}$ \\ ($\sim$~$1.37\times10^{2}$~)}     \\ \hline
			$^{14}$C single & $1.43\times10^{7}$ & $3.86\times10^{7}$ & $7.16\times10^{7}$ & $1.43\times10^{8}$ \\ \hline
			$^{14}$C double & \makecell[c]{$2.38\times10^{4}$ \\ ($\sim$~$2.38\times10^{3}$~)} & \makecell[c]{$1.73\times10^{5}$ \\ ($\sim$~$1.73\times10^{4}$~)} & \makecell[c]{$5.94\times10^{5}$ \\ ($\sim$~$5.94\times10^{4}$~)}& \makecell[c]{$2.38\times10^{6}$ \\ ($\sim$~$2.38\times10^{5}$~)} \\ \hline
			$^{14}$C triple & $1.97\times10^{1}$ & $3.88\times10^{2}$ & $2.47\times10^{3}$ & $1.97\times10^{4}$\\ \hline
			\makecell[c]{Signal-to-background\\ ratio:  ($\frac{\textit{pp}-\nu}{^{14}C double}$)} & \makecell[c]{$\sim$~1 : 17 \\} & \makecell[c]{$\sim$~1 : 126 \\} & \makecell[c]{$\sim$~1 : 431   \\  } & \makecell[c]{$\sim$~1: 1727   \\   }  \\ \hline 
		\end{tabular}
	\end{table*}
    
    For an LS detector whose sensitive target mass is $m$ kiloton (kton), the frequency of $^{14}$C single event is: 
    
    \begin{large} 
		\begin{equation} 
		f_{single} [Hz] = \frac{C_{^{14}C} \cdot N_{A} \cdot m}{\tau \cdot M} \times 10^{9}
		\label{eq:frequency_single}
		\end{equation} 
	\end{large}
    
    where $N_{A}$ is Avogadro's constants ($6.023 \times 10^{23}$), and $\tau$, $M$, $C_{^{14}C}$ correspond to $^{14}$C's lifetime, molar mass and its concentration in LS, respectively. 
    
    The frequency of $^{14}$C pile-up events can be calculated as follows:
    
     \begin{large} 
		\begin{equation} 
		f_{pile-up} [Hz] = \frac{e^{-f_{single} \cdot \Delta t}}{(n-1)!} \cdot f^{n}_{single} \cdot \Delta t^{n-1} \cdot \varepsilon
		\label{eq:frequency_pile-up}
		\end{equation} 
	\end{large}
    
    Where $n$ ($n \geq 2$) denotes the multiplicity of $^{14}$C accidental coincidence, for example, $n = 2$ represents the case of double $^{14}$C pile-up. $\Delta t$ is the time window for detection and $\varepsilon$ corresponds to the reconstruction efficiency of $^{14}$C pile-up events. 
    
\begin{figure}[!htb]
		\centering
		\includegraphics[width=1\hsize]{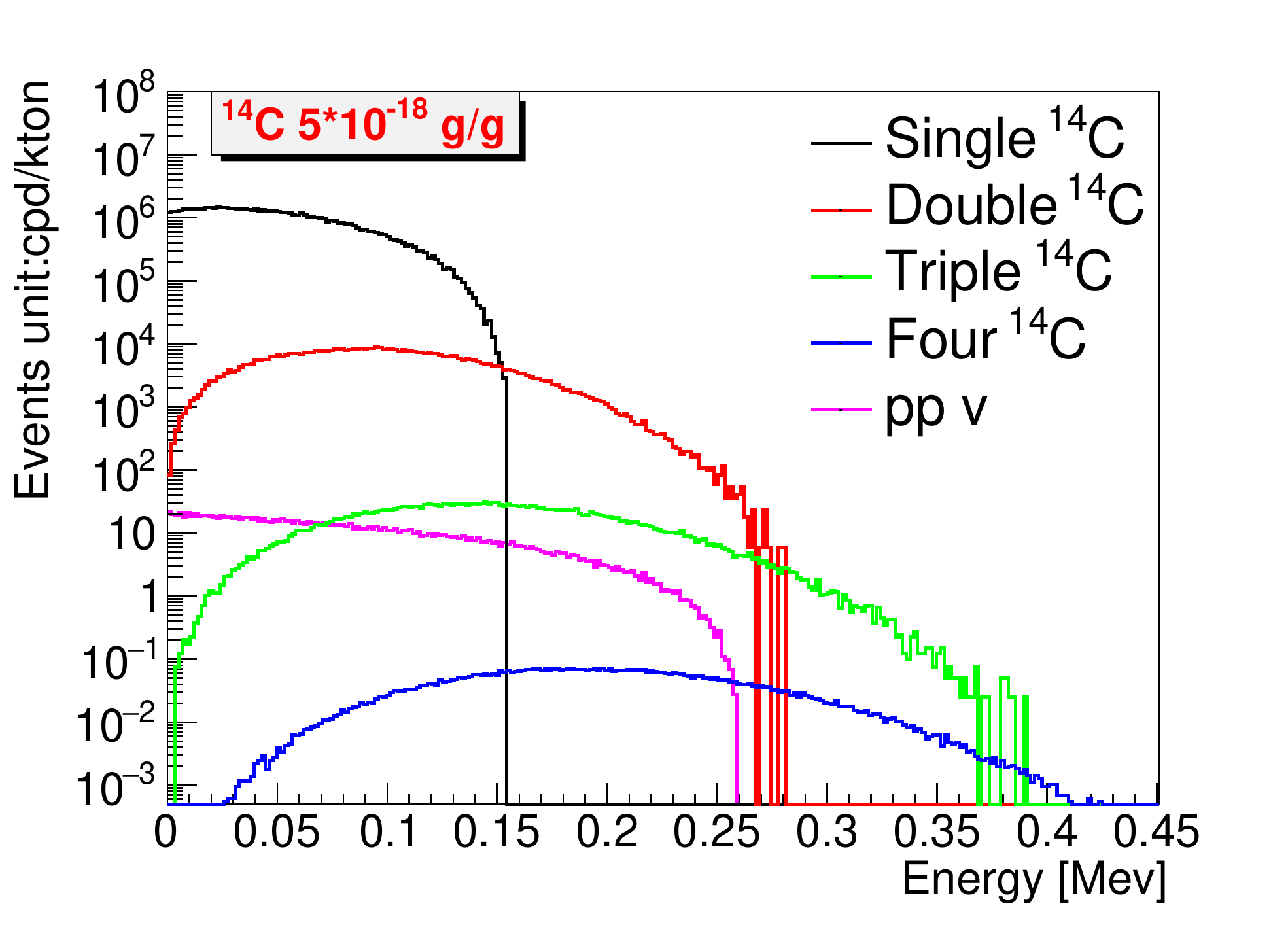}
		\caption{The recoil energy spectra of \textit{pp} neutrinos, $^{14}$C single, double, and triple pile-up events in a spherical LS detector, whose radius and $^{14}$C concentration are 15~m and $5 \times 10^{-18}$~g/g, respectively. The spectra do not include the detection effects: energy non-linearity, non-uniformity, and resolution. The contribution from $^{14}$C pile-up with higher order is negligible and not shown.} 
		\label{fig:spectrum_compare} 
	\end{figure}

    As the detector mass increases, the dramatic increase in $^{14}$C pile-up events has to be taken into account and rejected effectively. Taking a large spherical LS detector as an example, assuming the radius of the detector is 15~m thus the detector mass is about 12~kton, Table~\ref{table:tab-eventRate} lists the event rate of \textit{pp} neutrinos, $^{14}$C single and pile-up events in different $^{14}$C concentrations. A 500~ns time window was used in this calculation and reconstruction efficiency was set to 100\%. Assuming the $^{14}$C concentration of LS is $5 \times 10^{-18}$~g/g in the above detector, Fig.~\ref{fig:spectrum_compare} shows the recoil energy spectrum of \textit{pp} neutrinos via elastic neutrino-electron scattering, its calculation can be found in~\cite{Xu-2022wcq}. The energy spectra of $^{14}$C single, double, and triple pile-up events are shown for comparison. For this giant detector, \textit{pp} neutrino signals are totally swamped by $^{14}$C pile-up events of more than two orders of magnitude.
    
    In Table~\ref{table:tab-eventRate}, the values in brackets indicate the event rates inside the energy range of interest, which is from 0.16~MeV to 0.25~MeV for deposited energy by considering that the \textit{Q} value of $^{14}$C $\beta$ decay is $\sim156$~keV and the scattered electron of \textit{pp} neutrino is difficult to distinguish from the emitted electron of $^{14}$C single event. The target mass of the above detector ($\sim$12~kton) is $\sim$43 times larger compared to Borexino ($\sim$278~ton). As a result, the signal-to-background ratio of $pp$ neutrinos and $^{14}$C double pile-up events is smaller than 1 : 126 in the case of $^{14}$C concentration at $2.7 \times 10^{-18}$~g/g in this detector, and the signal-to-background ratio will be much poorer if an unlucky $^{14}$C concentration was found. On the other hand, since the energy resolution will introduce smearing in the energy spectrum, the energy range of analysis needs to be determined based on realistic situations.  
    
    More neutrino experiments are under construction or being planned, many of them~\cite{JUNO-PPNP, Jinping-2016iiq, DARWIN-2020bnc, Bieger-2021sas, juno-yellowbook,JUNO-PPNP, LENA} have good potential in \textit{pp} neutrino detection since they are expected to have a large detector target, well-controlled radioactivity, low detection threshold or good energy resolution. For those experiments with LS detectors of tens of kilotons, $^{14}$C pile-up makes the detection difficult at the low energy region. Hence it is necessary to develop an approach for $^{14}$C pile-up discrimination and reduction, especially the discrimination of $^{14}$C double pile-up since its event rate is much higher than other accidental coincidences. 
     
    This paper focuses on the discrimination of \textit{pp} solar neutrinos and $^{14}$C double pile-up events. As for the discrimination of other accidental coincidences with $^{14}$C multiplicity $\geq$ 3, it is an important topic in the case of poor $^{14}$C concentration, but it is not the subject of this article. The details of our work will be presented as follows: First, we build an LS detector in simulation and investigate the features of the detector's PMT hit pattern for \textit{pp} neutrinos and $^{14}$C double pile-up events (Sec.~\ref{section:Simulation}). Then, we present several approaches for $^{14}$C double pile-up discrimination based on multivariate analysis and deep learning technology (Sec.~\ref{section:Discrimination methods}). In Sec.~\ref{section:Discrimination performance and discussion}, the discrimination performance will be shown and compared. Finally, a summary will be provided in Sec.~\ref{section:Summary}
    
    \begin{figure}[!htb]
        \flushleft
		\centering
		\includegraphics[width=1.\hsize]{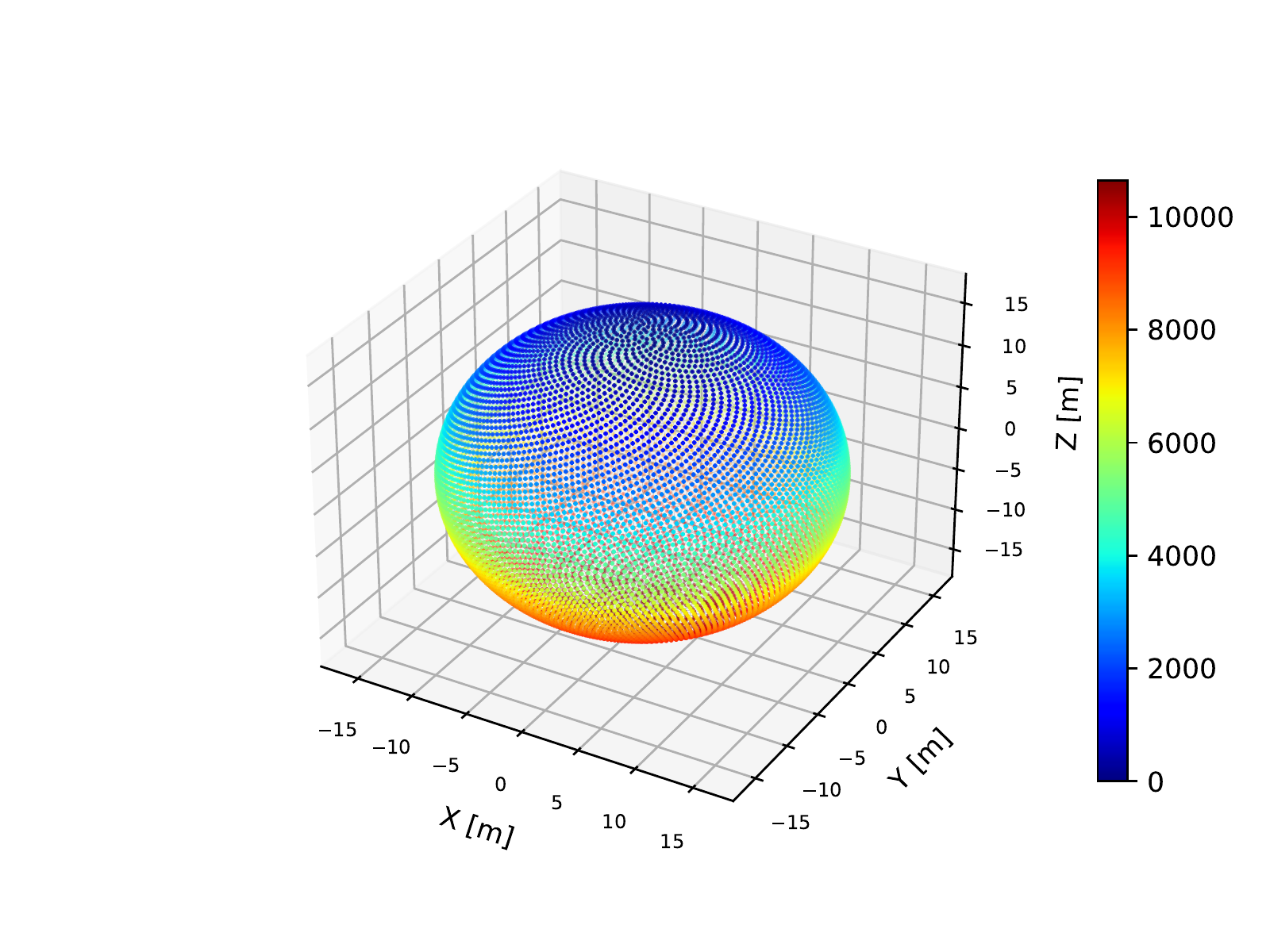}
		\caption{A schematic view of the detector. Each pixel corresponds to a 20-inch PMT, and its color indicates the ID of each PMT. In total, there are 10650 PMTs.} 
		\label{fig:detector} 
	\end{figure}

\section{Detector simulation}
\label{section:Simulation}

    In this work, a spherical LS detector was built in Monte Carlo (MC) simulation using Geant4 toolkit~\cite{GEANT4-2002zbu}, version 4.10.p02. The radius of the spherical detector is 15~m, and the LS is contained in an acrylic sphere with 10~cm thick. To simplify the simulation, a sensitive optical surface is defined for photons receiving instead of the PMT simulation in detail. The sensitive optical surface is a sphere outside the acrylic sphere, separated by 1~m thick water. Next, the coverage and the quantum efficiency of the photosensors can be easily applied and tuned. In the simulation, the coverage rate is 65\% and it corresponds to about 10650 20-inch photomultipliers (PMTs) uniformly distributed on the sensitive optical surface. Fig.~\ref{fig:detector} shows the schematic view of the detector. In the simulation, a 30\% averaged quantum efficiency was used for 20-inch PMTs with 2\% Gaussian relative spread. The LS properties were referenced from~\cite{Zhou-2015gwa, Gao-2013pua, Wurm-2010ad, Zhang-2020mqz, Ding-2015sys, Buck-2015jxa, OKeeffe-2011dex, Yu-2022god}, and comprehensive optical processes were adopted, including quenching, Rayleigh scattering, absorption, and re-emission. Table~\ref{table:tab-SimParameters} summarizes the main parameters of PMTs in the simulation, including the transit time spread (TTS), quantum efficiency (QE), dark noise, and the resolution of single photoelectron (spe). As a result, about 1100 photoelectrons (PEs) will be observed by 10650 PMTs for a 1 MeV electron fully deposited its kinetic energy in the center of the detector, and it corresponds to about 3\% energy resolution. On the other hand, there are about 105 additional PEs that will be detected, which come from the PMT dark noise in a time window of 500~ns.  

    \begin{table}
	\caption{PMT parameters in the simulation.}
	\label{table:tab-SimParameters}
		\renewcommand\arraystretch{1.4}
		\setlength{\tabcolsep}{7mm}
		\centering
		\begin{tabular}{c|c}
			\hline
			 Parameters & Values   \\ \hline
			 PMT Coverage & 65\%   \\ 
			 PMT QE & 30\% $\pm$ 2\% (Gaussian)  \\ 
			 PMT TTS & 3 $\pm$ 0.3~ns (Gaussian)  \\ 
			 PMT dark rate & 20 $\pm$ 3~kHz (Gaussian) \\ 
			 PMT spe resolution & 30\% $\pm$ 3\% (Gaussian)  \\ 
			 Time window & 500~ns   \\ \hline
			 \end{tabular}
    \end{table}

    \begin{figure*}[!htb]
        \centering
        \subfigure[]{
            \includegraphics[width=0.45\hsize]{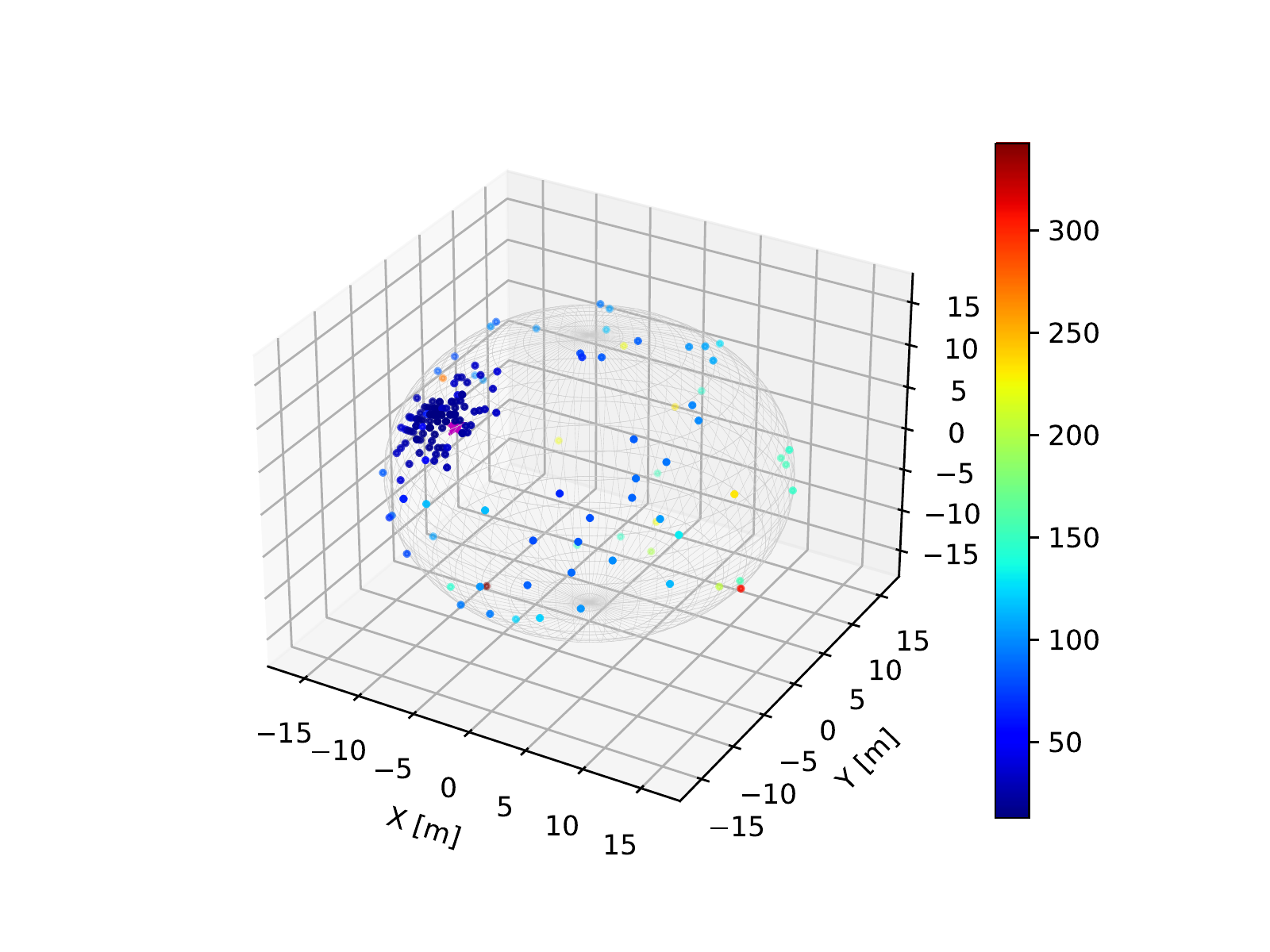}
            \label{fig:MCsample_pp}
        }
        \quad
        \subfigure[]{
            \includegraphics[width=0.45\hsize]{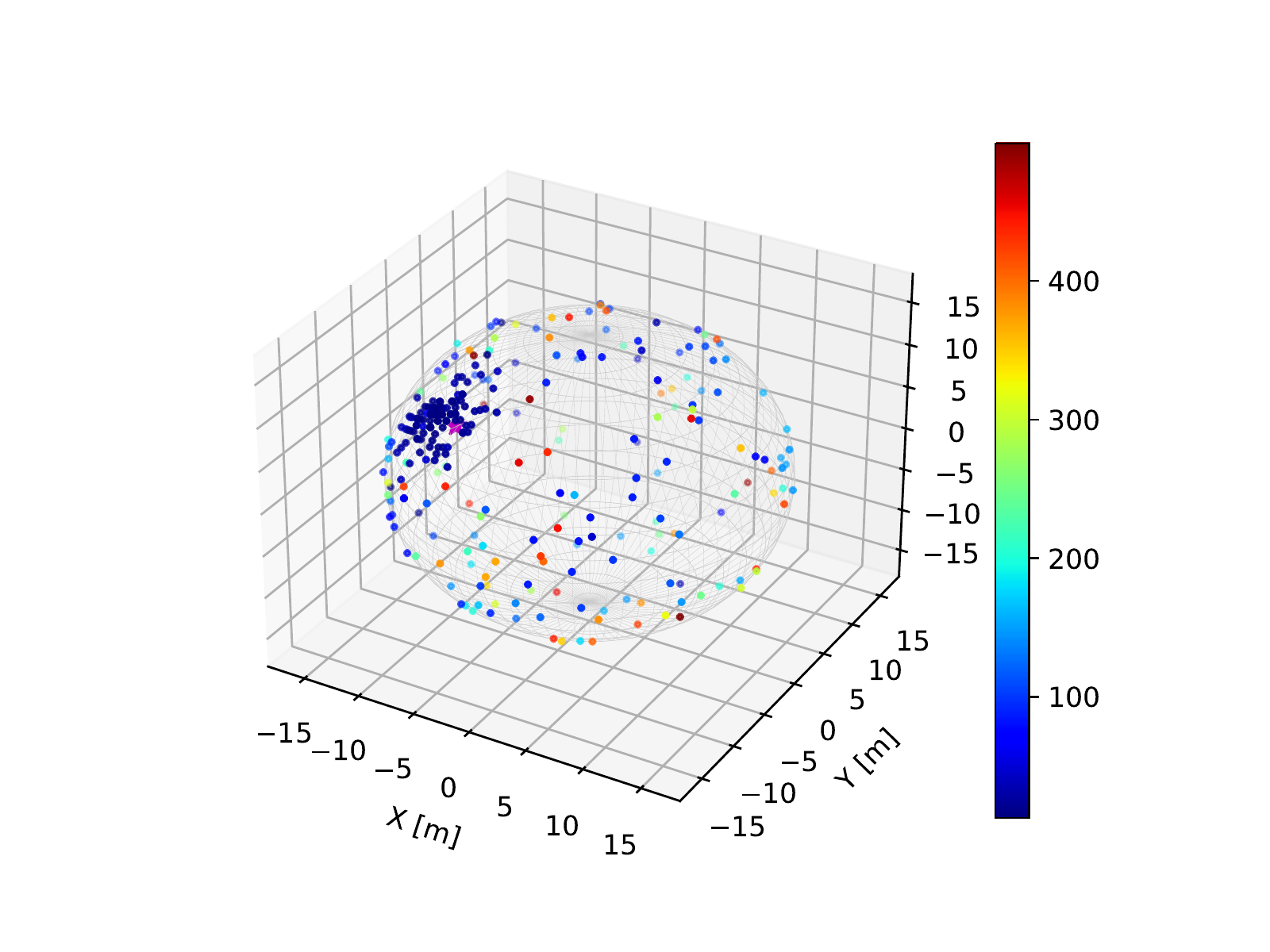}
            \label{fig:MCsample_pp_DR}
        }
        \caption{The PMT hit patterns of a \textit{pp} solar neutrino event. Each pixel corresponds to a fired PMT and its color indicates the hit time information. The location of the purple star is (-6582.21, -8972.86, 8696.34)~mm, which indicates the position where the physics event deposited its energy (159.94~keV). (a) only physics hits are included, and 172 PEs are observed for a 500~ns time window. (b) Both physics hits and PMT dark noise hits are shown, and 284 PEs are observed for a 500~ns time window, including 112 PEs from PMT dark noise.}
        \label{fig:hitpatterns_pp}
    \end{figure*}

    \begin{figure*}[!htb]
        \centering
        \subfigure[]{
            \includegraphics[width=0.45\hsize]{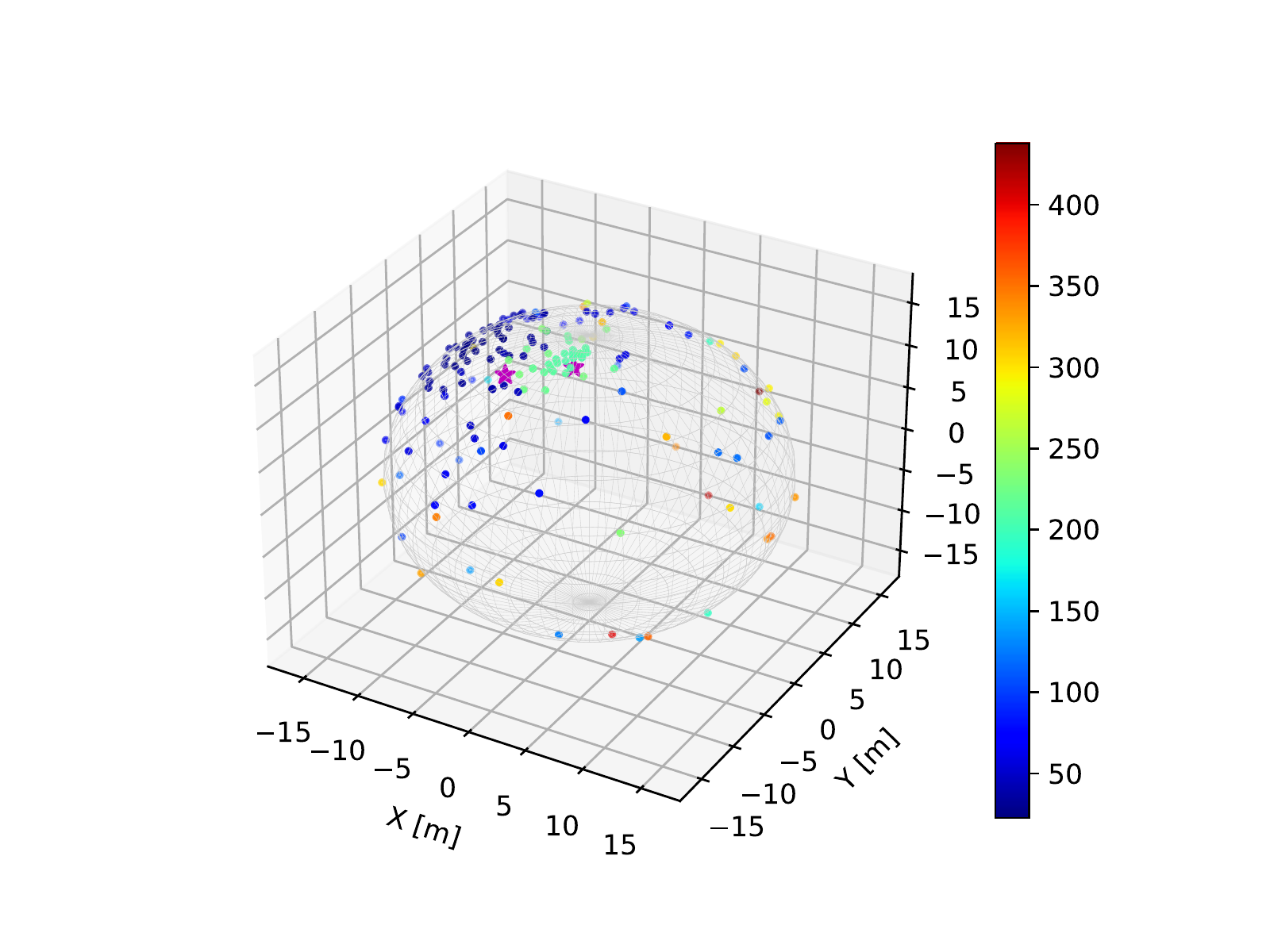}
            \label{fig:MCsample_C14_double}
        }
        \quad
        \subfigure[]{
            \includegraphics[width=0.45\hsize]{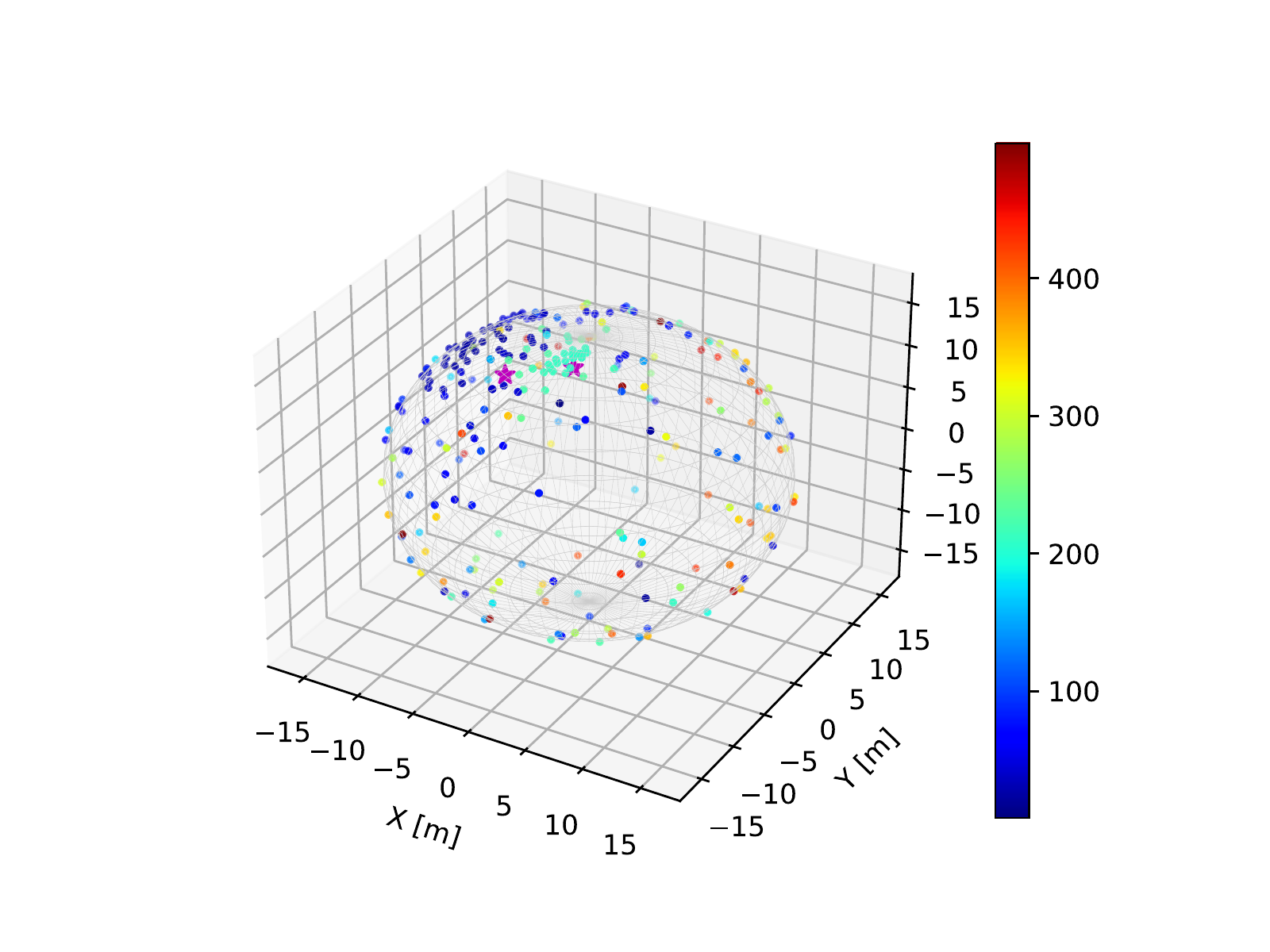}
            \label{fig:MCsample_C14_double_DR}
        }
        \caption{The PMT hit patterns of a $^{14}$C double pile-up event. Each pixel corresponds to a fired PMT and its color indicates the hit time information. Two purple stars indicate the positions where two $^{14}$C events deposited their energies (71.161~keV and 56.593~keV). Their locations are (-6229.32, -2139.36, 10471.7)~mm and (484.61, -3199.44, 14423.5)~mm, respectively. (a) only physics hits are included, and 173 PEs (107+66) are observed for a 500~ns time window. (b) Both physics hits and PMT dark noise hits are shown, and 273 PEs are observed for a 500~ns time window, including 100 PEs from PMT dark noise. }
        \label{fig:hitpatterns_C14_double}
    \end{figure*}

    To investigate the response features of \textit{pp} neutrinos and $^{14}$C double pile-up events, their MC samples were generated and compared. About 1 million final-state electrons from the elastic neutrino-electron scattering reaction of \textit{pp} neutrinos were uniformly simulated in the LS volume, and the spectrum of scattered electrons was referenced from~\cite{Xu-2022wcq}. Since the final-state electrons from the elastic neutrino-electron scattering are similar to the emitted electrons from $^{14}$C $\beta$ decay ($^{14}$C single event), it is difficult to distinguish them in event-by-event level. Therefore, an energy cut is needed to focus on a narrow energy region. The same treatment is applied by Borexino. On the other hand, there is about 5\% energy non-linearity~\cite{Yu-2022god, DayaBay-2019fje} for electrons whose kinetic energy is around 0.2~MeV in LS, and energy resolution is included in the above simulation naturally. As a result, in our analysis, a 255~PEs cut was applied to the total number of photoelectrons of all PMTs by considering $\sim$156~keV end-point energy of $^{14}$C $\beta$ decay ($\sim$150~PEs) and the contribution of PMT dark noise ($\sim$105~PEs). 
    
    After the total PEs cut, an MC sample which includes 100 thousand of \textit{pp} neutrino will be used for the study of discrimination, and they are uniformly distributed in the LS. As for the generation of $^{14}$C double pile-up sample, firstly, a large dataset was produced by simulating 10 million $^{14}$C single events in the LS via $^{14}$C $\beta$ decay. Next, two $^{14}$C single events were randomly picked up from the dataset and then merged into a double pile-up event. In the merge operation, since the lifetime of $^{14}$C is longer than 8000 years, the time interval of two $^{14}$C single events can be approximately treated as a uniform distribution in a few hundred nanoseconds. Similarly, a 255~PEs cut was applied and 100 thousand $^{14}$C double pile-up events will be used for our analysis.

    As illustrated in Fig.~\ref{fig:hitpatterns_pp} and Fig.~\ref{fig:hitpatterns_C14_double}, \textit{pp} solar neutrinos and $^{14}$C double pile-up events show different features in their time and spatial distributions. \textit{pp} solar neutrino is a single point-like event whose energy deposition occurs in a relatively short time and a small space; hence, only one cluster will be found in its PMT hit pattern. As for $^{14}$C double pile-up event, if two $^{14}$C decay at different detector positions, two clusters will be found. On the other hand, since the hit time distribution of the fired PMTs includes both scintillation time and photon's time of flight, as well as the decay time of $^{14}$C, this makes the hit time distribution useful for identification studies. Especially for the case that two $^{14}$C decay near each other, their spatial distribution will not be significantly different from a single point-like event, but the hit time distribution may still be helpful if the time interval between two $^{14}$C decays is large. An example of this case can be found in Fig.~\ref{fig:hitpatterns_C14_double}. In Sec.~\ref{section:Discrimination methods}, event spatial information will be extracted and used together with hit time information as input to the discrimination algorithms.
    
\section {Discrimination methods}
\label{section:Discrimination methods}

    The basic idea to develop a discrimination algorithm for \textit{pp} solar neutrinos and $^{14}$C double pile-up events is by utilizing their time and spatial information, which have different characteristics (Sec.~\ref{section:Simulation}). Similar approaches were applied in the discrimination of single-site and multi-site energy depositions in large-scale liquid scintillation detectors~\cite{Dunger:2019dfo}. During the measurement, the cluster structure will be smeared by interference from the dark noise and the TTS of the PMT. These effects make the identification becomes more challenging and requires efficient approaches. In this study, a multivariate analysis using the TMVA (Toolkit for Multivariate Data Analysis)~\cite{Hocker:2007ht,Speckmayer:2010zz} is performed, and the widely used algorithm BDTG (Boosted Decision Trees with Gradient boosting) is chosen, and investigated. In addition, deep learning technologies based on the VGG network have also been applied. In the following, we present the details of the discrimination method. 
    
\subsection{TMVA analysis}
\label{subsection: TMVA analysis}

    \begin{figure*}[!htb]
        \centering
        \subfigure[]{
            \includegraphics[width=0.3\hsize]{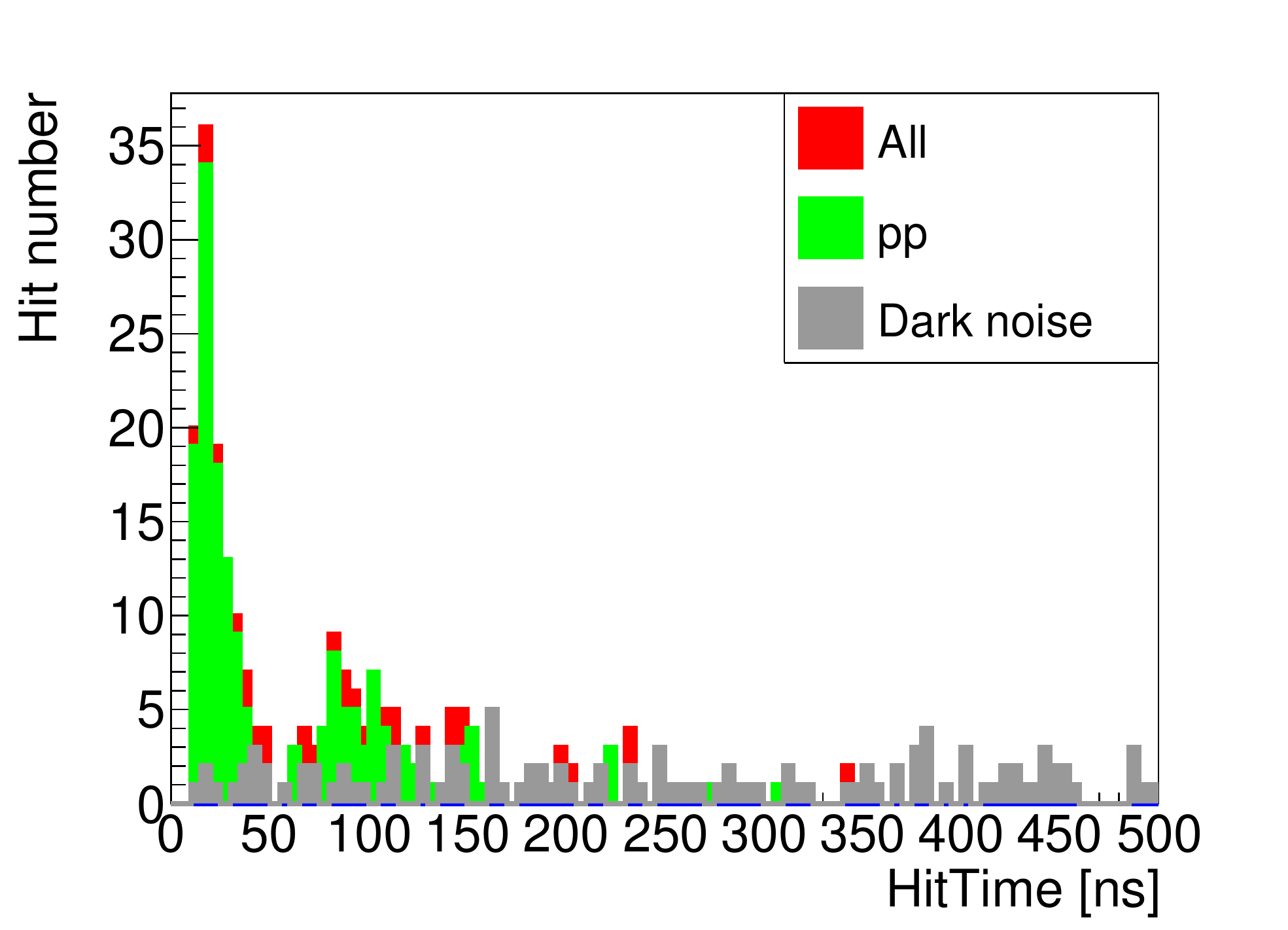}
            \label{fig:hittime_pp}
        }
        \quad
        \subfigure[]{
            \includegraphics[width=0.3\hsize]{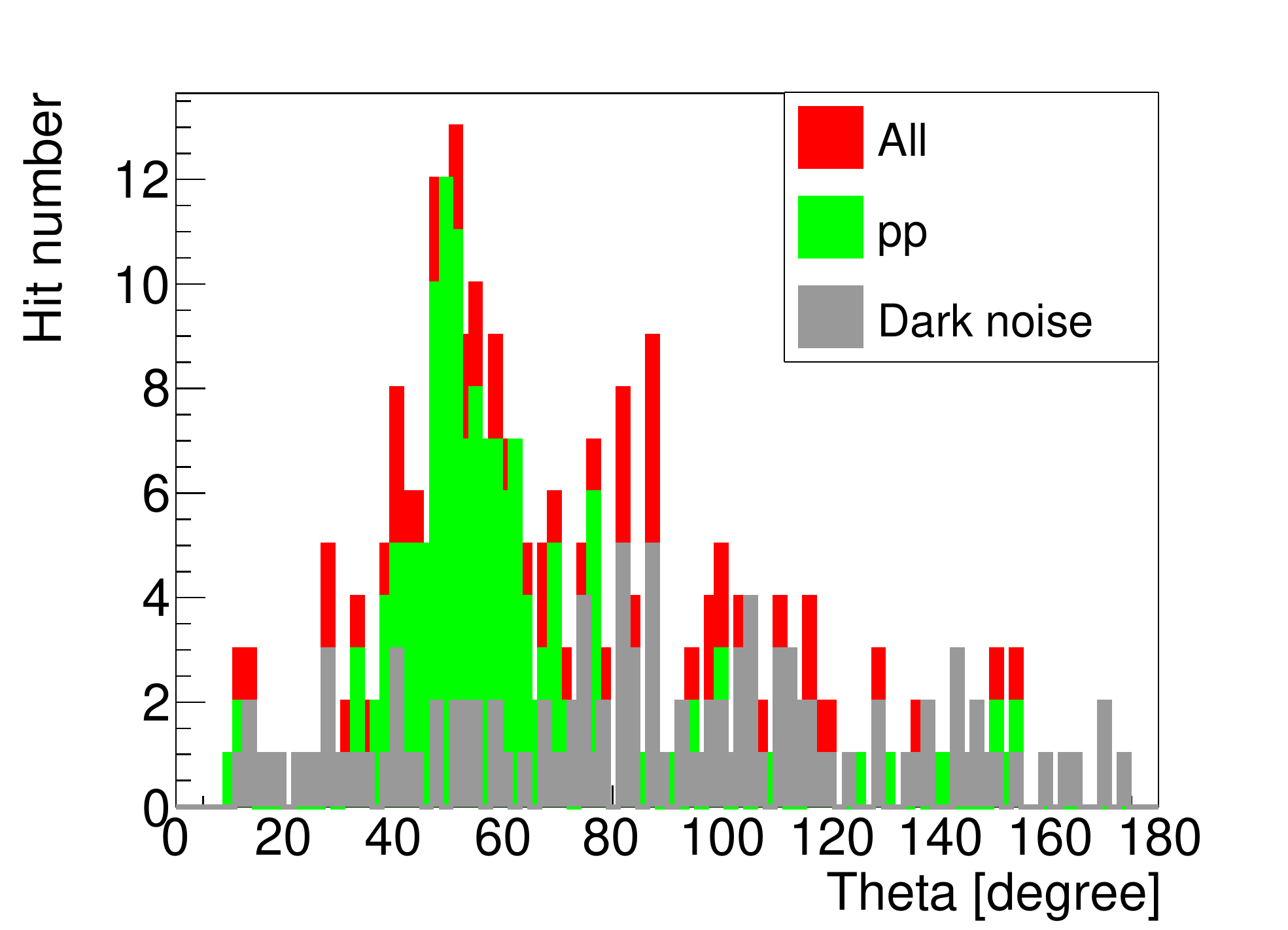}
            \label{fig:theta_pp}
        }
        \quad
        \subfigure[]{
            \includegraphics[width=0.3\hsize]{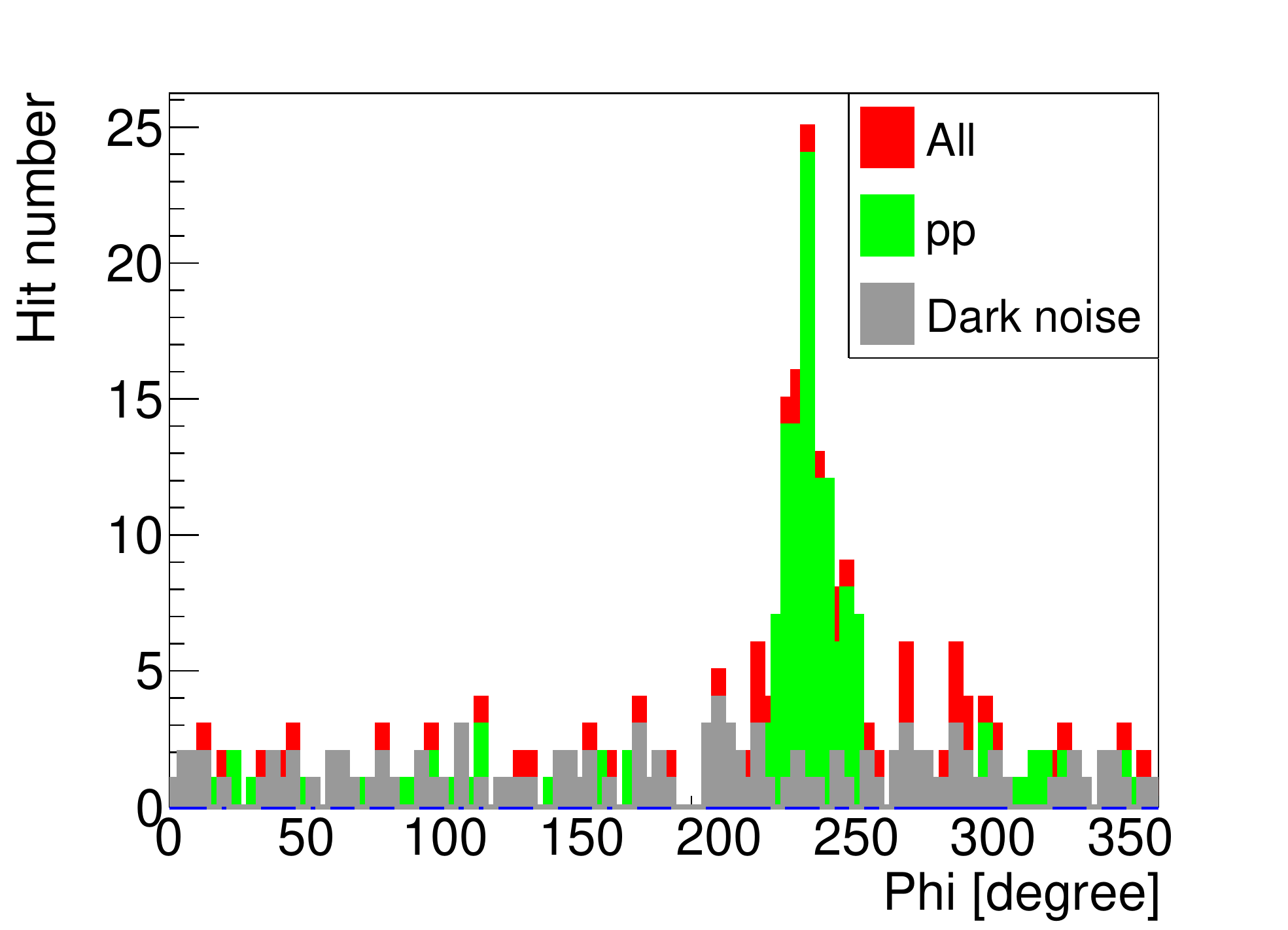}
            \label{fig:phi_pp}
        }
        \caption{The hit time, $\theta$, and $\phi$ distribution of a \textit{pp} solar neutrino event, which corresponds to the event in Fig.~\ref{fig:MCsample_pp_DR}. (a) hit time distribution. (b) $\theta$ distribution. (c) $\phi$ distribution. }
        \label{fig:distribution_pp}
    \end{figure*}

    \begin{figure*}[!htb]
        \centering
        \subfigure[]{
            \includegraphics[width=0.3\hsize]{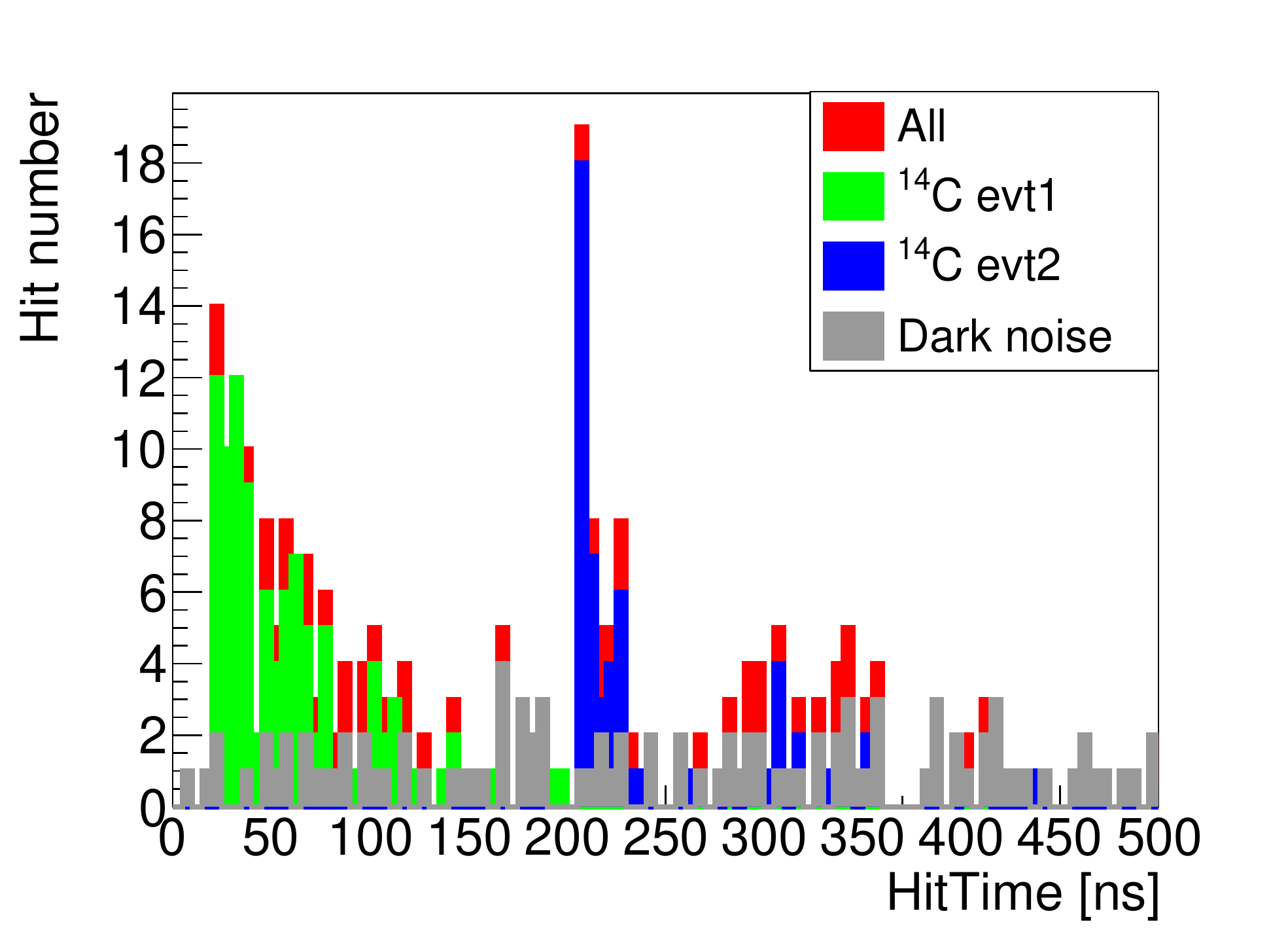}
            \label{fig:hittime_C14_double}
        }
        \quad
        \subfigure[]{
            \includegraphics[width=0.3\hsize]{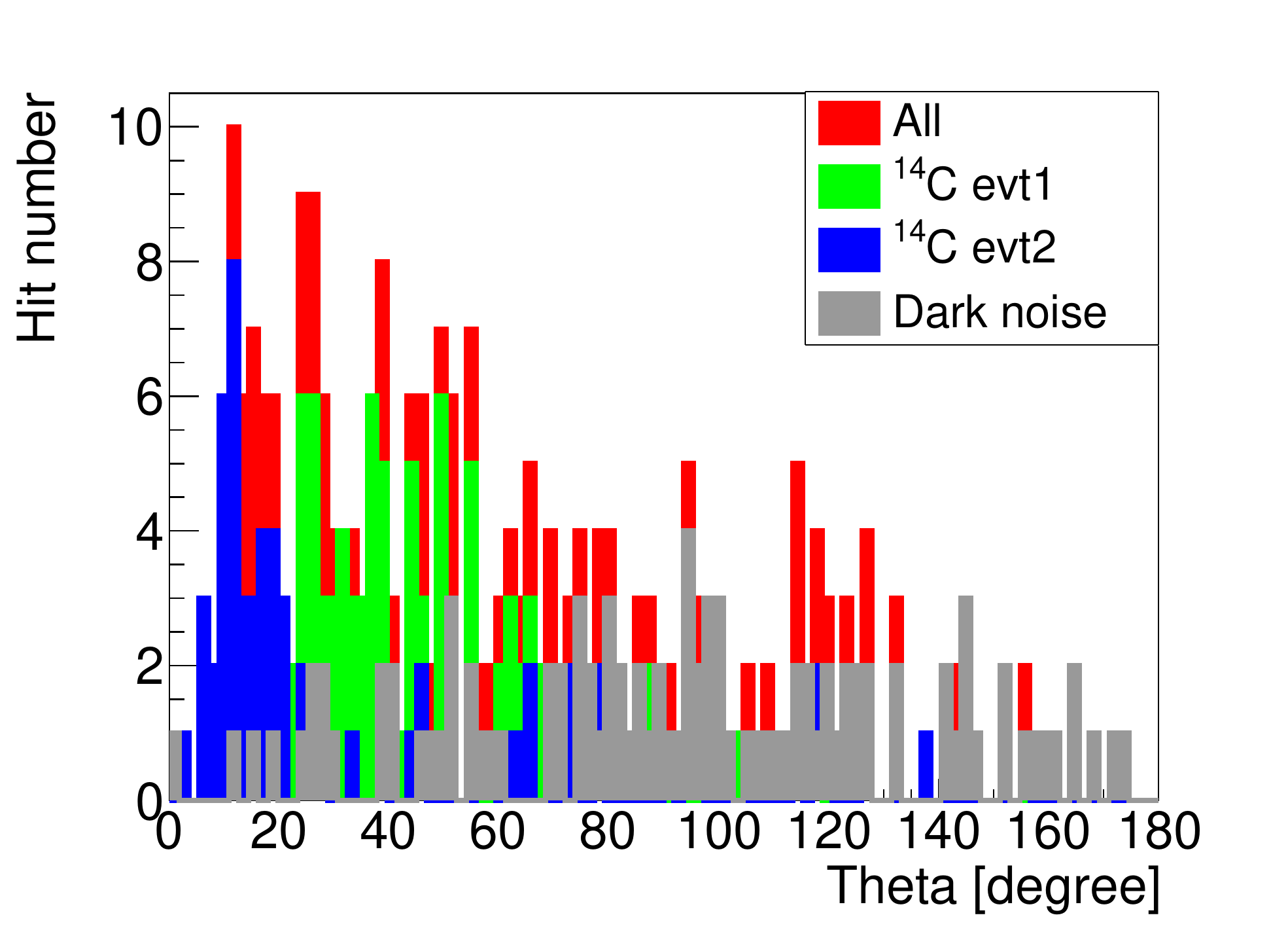}
            \label{fig:theta_C14_double}
        }
        \quad
        \subfigure[]{
            \includegraphics[width=0.3\hsize]{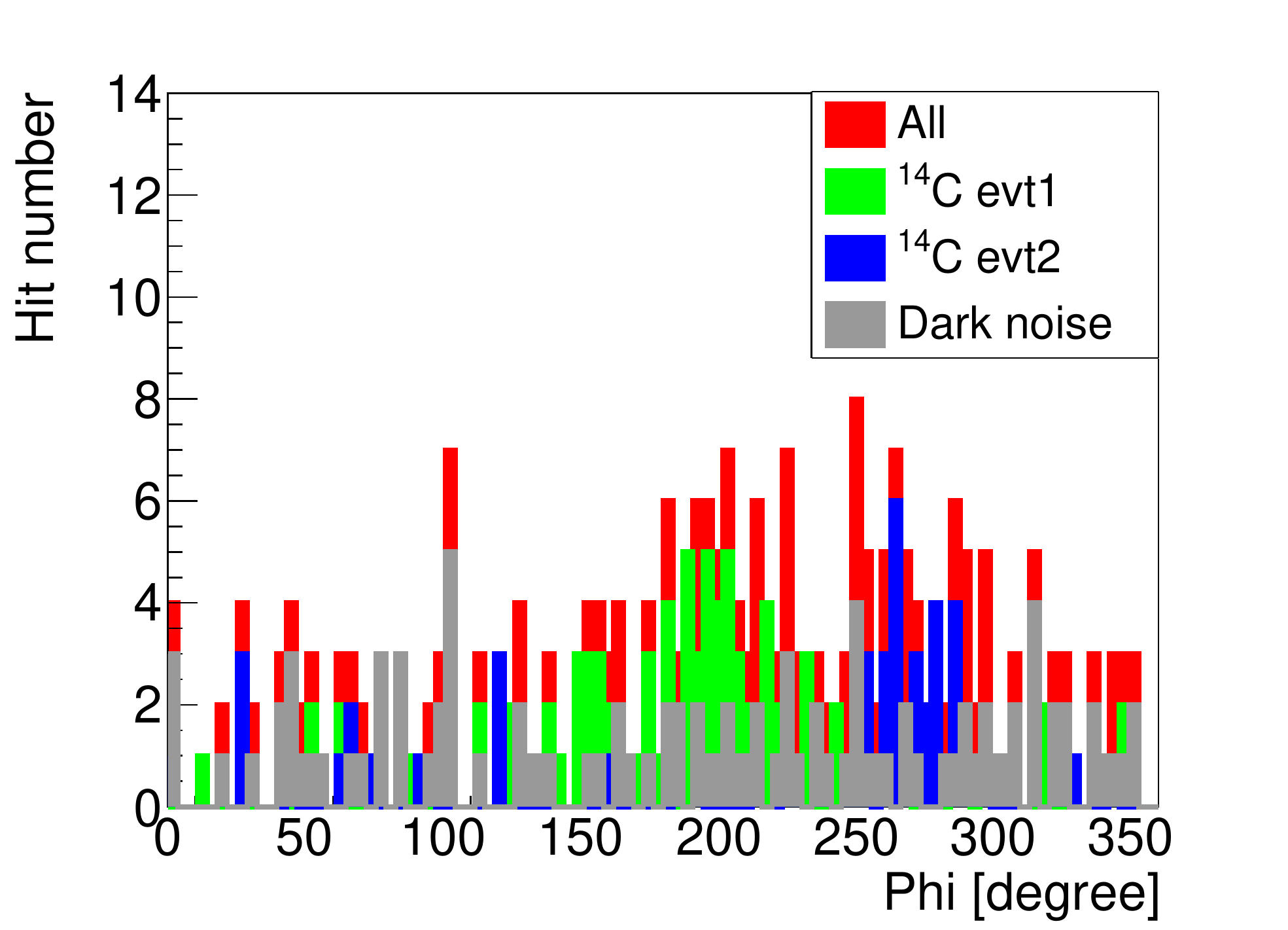}
            \label{fig:phi_C14_double}
        }
        \caption{The hit time, $\theta$ and $\phi$ distribution of a $^{14}$C double pile-up event, which corresponds to the event in Fig.~\ref{fig:MCsample_C14_double_DR}. (a) hit time distribution. (b) $\theta$ distribution. (c) $\phi$ distribution.}
        \label{fig:distribution_C14_double}
    \end{figure*}

    \begin{figure*}[!htb]
		\flushleft
		\centering
		\includegraphics[width=1.\hsize]{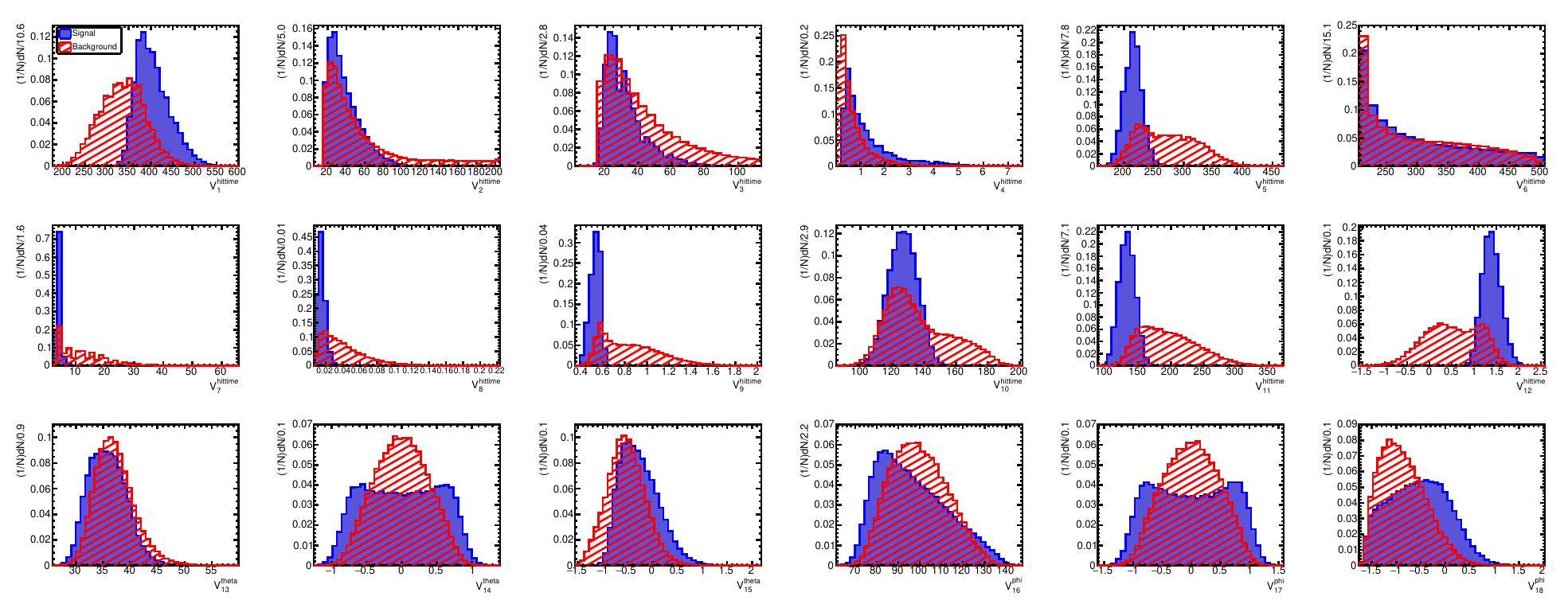}
		\caption{Normalized distributions of the variables of \textit{pp} solar neutrino and $^{14}$C double pileup event. } 
		\label{fig:TMVA_parameterInput_example} 
	\end{figure*}

    \begin{figure*}[!htb]
        \centering
        \subfigure[]{
            \includegraphics[width=0.4\hsize]{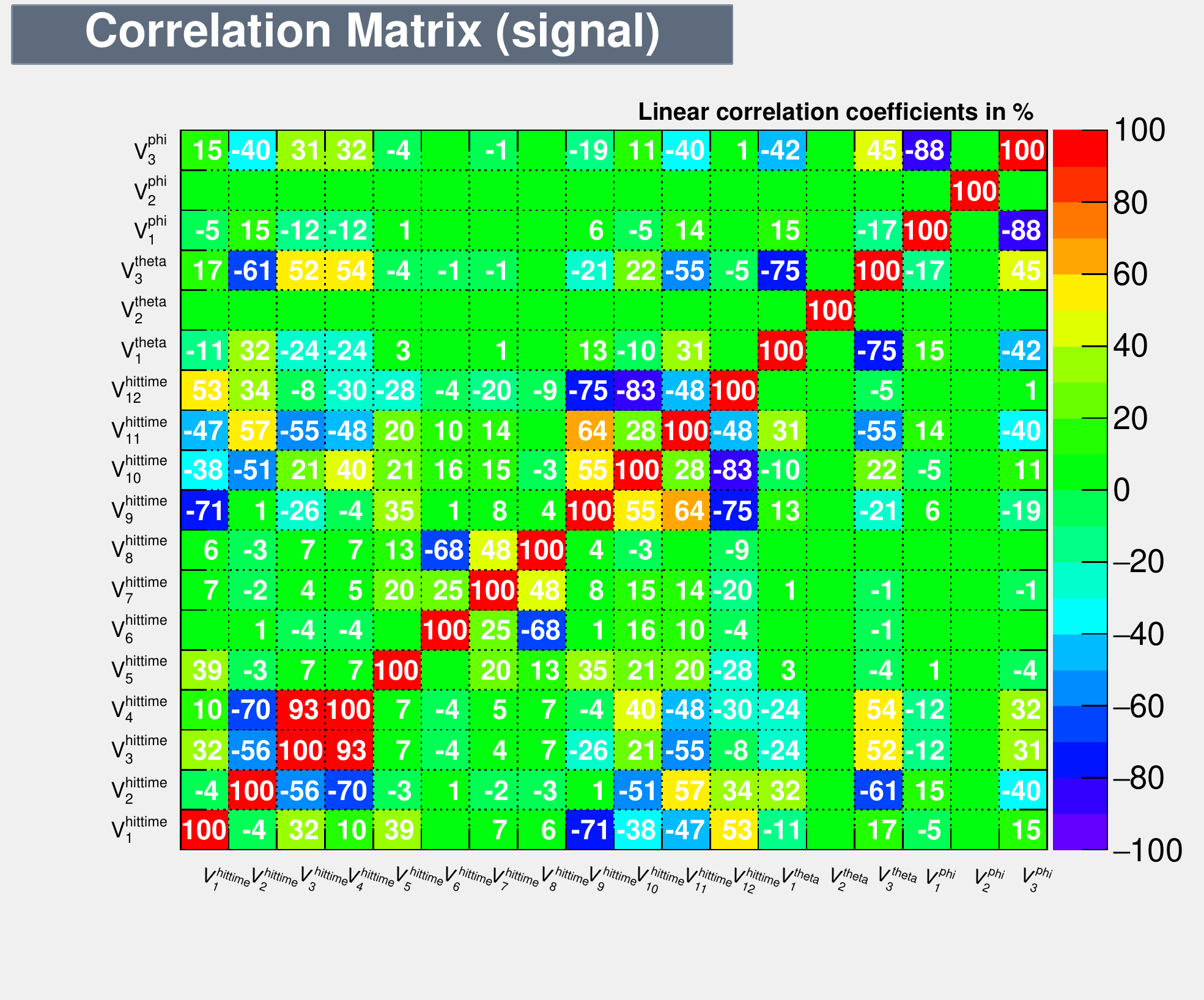}
            \label{fig:TMVA_parameterInput_correlation_signal}
        }
        \quad
        \subfigure[]{
            \includegraphics[width=0.4\hsize]{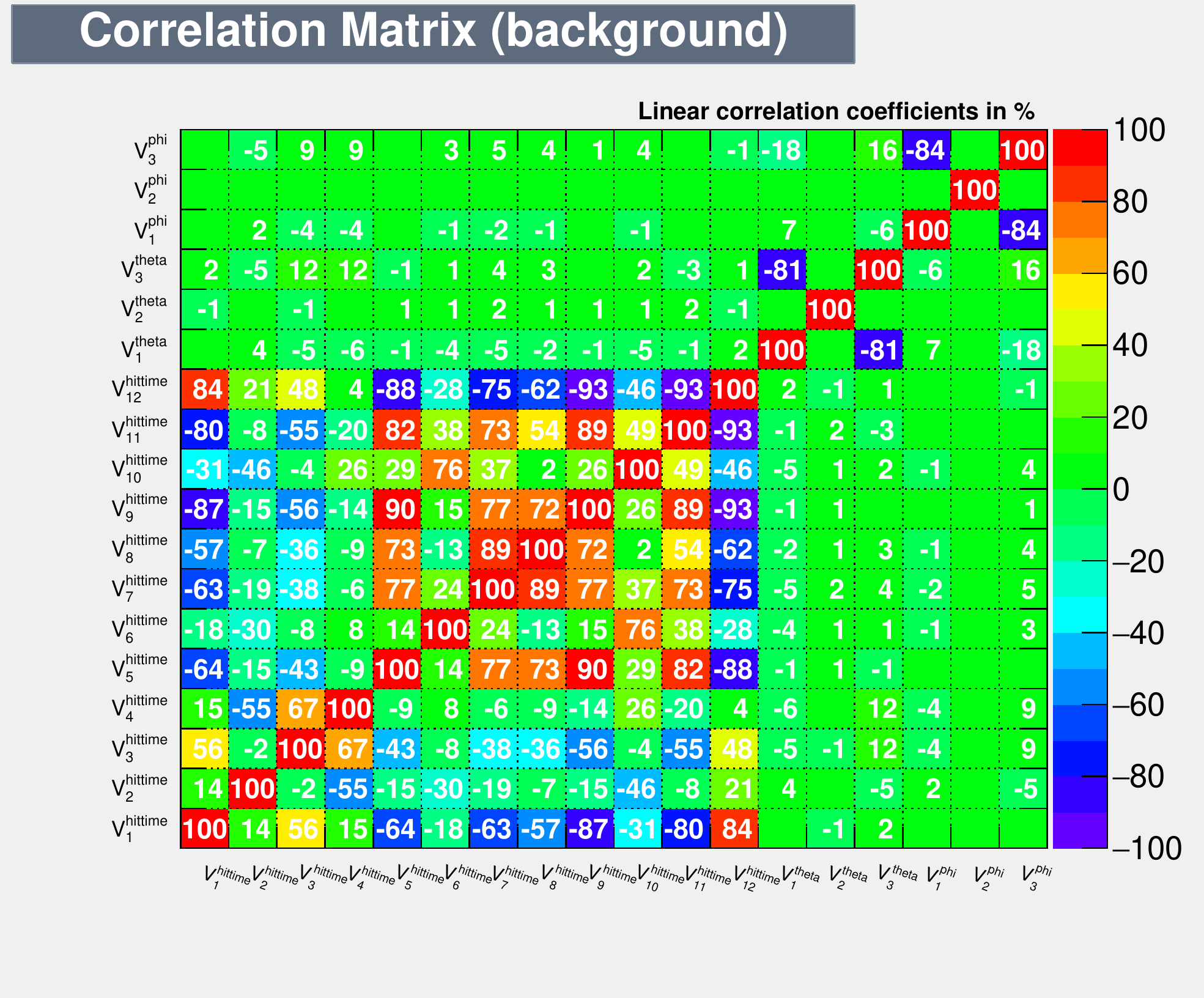}
            \label{fig:TMVA_parameterInput_correlation_bkg}
        }
        \caption{Linear correlation matrix for the input variables of \textit{pp} solar neutrinos (a) and $^{14}$C double pile-up events (b).}
        \label{fig:Correlation_matrix}
    \end{figure*}

    TMVA~\cite{Hocker:2007ht,Speckmayer:2010zz} is a powerful tool for multivariate analysis, and it has been successfully applied in both signal and background classification in accelerator physics~\cite{Lampen:2008zza}, component identification of cosmic rays~\cite{LHAASO:2019qdu} and event reconstruction in LS detectors for neutrino experiments~\cite{Qian-2021vnh}. The TMVA toolkit hosts a large variety of multivariate classification algorithms. In this paper, we choose and investigate the TMVA algorithm BDTG. To extract input variables, the PMT hit pattern was projected onto a one-dimensional (1-D) plane for hit time, and $\theta$ and $\phi$ of each fired PMT in spherical coordinates, respectively. The projection results of Fig.~\ref{fig:MCsample_pp_DR} are shown in Fig.~\ref{fig:distribution_pp} and the projection results of Fig.~\ref{fig:MCsample_C14_double_DR} are shown in Fig.~\ref{fig:distribution_C14_double}. The \textit{pp} solar neutrino, which is a single point-like event, only shows one cluster in its distributions, while the $^{14}$C double pile-up event shows two clusters. 
    
    These 1-D distributions will be used in the multivariate analysis. The input variables of TMVA algorithms should be sensitive to discrimination and contain the characteristics of \textit{pp} solar neutrinos and $^{14}$C double pile-up events. In our analysis, it was found that the hit time information dominates the discrimination performance, so more variables are extracted from the 1-D distribution of hit time. A total of eighteen variables were used in the TMVA analysis. These variables are marked as $V^{\alpha}_{i}$, where $i = 1, 2, 3$, etc., and they correspond to the extracted parameters in each 1-D distribution. $\alpha = hittime$, $\theta$ or $\phi$, which denotes that the variables are from the 1-D distribution of hit time, $\theta$ or $\phi$. Their details can be found in Table~\ref{table:tab-TMVAVariables}.
    
    \begin{table}
	\caption{Input variables for multivariate analysis.}
	\label{table:tab-TMVAVariables}
		\renewcommand\arraystretch{1.4}
		\setlength{\tabcolsep}{7mm}
		\begin{tabular}{p{0.5cm}<{\raggedright}|p{5cm}<{\raggedright}}
			\hline
			 Variable & Description   \\ \hline
			 $V^{hittime}_{1}$ & Number of hits in the first 200~ns \\ \hline
             $V^{hittime}_{2}$ & The peak position of the highest bin in the first 200~ns \\ \hline
             $V^{hittime}_{3}$ & The amplitude of the highest bin in the first 200~ns \\ \hline
             $V^{hittime}_{4}$ & The ratio between the peak amplitude and the peak position of the highest bin in the first 200~ns \\ \hline
             $V^{hittime}_{5}$ & Number of hits in (200, 500)~ns \\ \hline
             $V^{hittime}_{6}$ & The peak position of the highest bin in (200, 500)~ns \\ \hline
             $V^{hittime}_{7}$ & The amplitude of the highest bin in (200, 500)~ns \\ \hline
             $V^{hittime}_{8}$ & The ratio between the peak amplitude and the peak position of the highest bin in (200, 500)~ns \\ \hline
             $V^{hittime}_{9}$ & The ratio between the number of hits in the first 200~ns and in (200, 500)~ns \\ \hline
             $V^{hittime}_{10}$ & The RMS value of the 1-D distribution of hit time \\ \hline
             $V^{hittime}_{11}$ & The Mean value of the 1-D distribution of hit time \\ \hline
             $V^{hittime}_{12}$ & The skewness coefficient of the 1-D distribution of hit time \\ \hline
             $V^{theta}_{1}$ & The RMS value of the 1-D distribution of $\theta$ \\ \hline
             $V^{theta}_{2}$ & The skewness coefficient of the 1-D distribution of $\theta$ \\ \hline
             $V^{theta}_{3}$ & The kurtosis coefficient of the 1-D distribution of $\theta$ \\ \hline
             $V^{phi}_{1}$ & The RMS value of the 1-D distribution of $\phi$ \\ \hline
             $V^{phi}_{2}$ & The skewness coefficient of the 1-D distribution of $\phi$ \\ \hline
             $V^{phi}_{3}$ & The kurtosis coefficient of the 1-D distribution of $\phi$ \\ \hline
			 \end{tabular}
    \end{table}
    
    Fig.~\ref{fig:TMVA_parameterInput_example} shows the normalized distributions of these input variables, and the difference in their shapes is observed by comparing the two types of events. On the other hand, the correlations of the input variables are checked for both \textit{pp} solar neutrinos and $^{14}$C double pile-up events. As shown in Fig.~\ref{fig:Correlation_matrix}, since we have dropped several variables with strong correlations in the previous study, the correlation of the current variables is acceptable with no one greater than 95\%.

    \begin{table*}[!htb]
	\caption{Parameters used in the BDTG algorithm.}
	\label{table:tab-BDTG-setting}
		\renewcommand\arraystretch{1.4}
		\setlength{\tabcolsep}{7mm}
		\centering
		\begin{tabular}{c|c|c}
			\hline
			 Configuration option & Setting & Description \\ \hline
			NTrees & 1000 & Number of trees in the forest     \\ 
			MaxDepth & 2 & Max depth of the decision tree allowed     \\ 
			MinNodeSize & 2.5\% & Minimum percentage of training events required in a leaf node     \\ 
			nCuts & 20 & Number of grid points in variable range used in finding optimal cut in node splitting     \\ 
			BoostType & Grad & Boosting type for the trees in the forest     \\ \hline
		\end{tabular}
	\end{table*}
	
    The MC samples of \textit{pp} solar neutrinos and $^{14}$C double pile-up events are divided into two equal parts, respectively, one for TMVA training and the other for validation. To improve the performance, several main parameters are tuned in the BDTG algorithm, Table~\ref{table:tab-BDTG-setting} shows the settings of the parameters, the other parameters are set to their default values and aren't listed in the tables.
   
\subsection{Deep learning}
\label{subsection: Machine learning}

    Deep learning technology is widely used in high energy physics and nuclear physics, there are many successful applications~\cite{Guest-2016iqz,Guest-2018yhq,He-2018nst,Ma-2019nst,Qian-2021vnh,Li-2022nst,Liu-2022nst}, such as energy reconstruction, track reconstruction, particle identification, signal processing, etc. In this paper, the deep learning algorithm VGG convolutional neural network is used for feature recognition of one-dimensional sequences. The extracted PMT hit patterns are projected into a one-dimensional feature series for hit time, and $\theta$, $\phi$, respectively, which is similar to Fig.~\ref{fig:distribution_pp} and Fig.~\ref{fig:distribution_C14_double}. To extract their features, a one-dimensional convolution kernel is used for the above three series, a pooling layer is used for information compression, and a fully connected layer is used for particle classification. The model structure is based on the architecture of VGG-16, which includes 13 convolution and pooling modules, 3 fully connected layers, batch normalization layers, and connected neural unit dropout processing.

    On the other hand, in addition to one-dimensional projection using the PMT hit patterns, we also tried two-dimensional projection methods to provide input to deep learning network, including  Mercator projection, sinusoidal projection, and the projection method based on the arrangement of PMTs~\cite{Qian-2021vnh}. However, after applying the two-dimensional projection, it is found that the performance does not improve but slightly decreases. Considering that the number of hits is very small in the energy range of interest, we performed a detailed investigation and comparison, and this result can be explained by the fact that the cluster features are much more pronounced in the one-dimensional projection, but they are very discrete in the two-dimensional projection.
    
    Finally, a one-dimensional projection is used to provide input to the VGG network described above. We trained the VGG network using Adaptive Momentum with a batch size of 256 samples, a momentum of 0.9, and an initial learning rate of 0.01. For every 10 epochs, the learning rate is reduced by a factor of 10. The accuracy of the model is evaluated using a cross-entropy loss function.

\section {Discrimination performance and discussion} 
\label{section:Discrimination performance and discussion}

\subsection{Discrimination performance of the BDTG model}
\label{subsection:result_BDTG}

    \begin{figure*}[!htb]
        \centering
        \subfigure[]{
            \includegraphics[width=0.4\hsize]{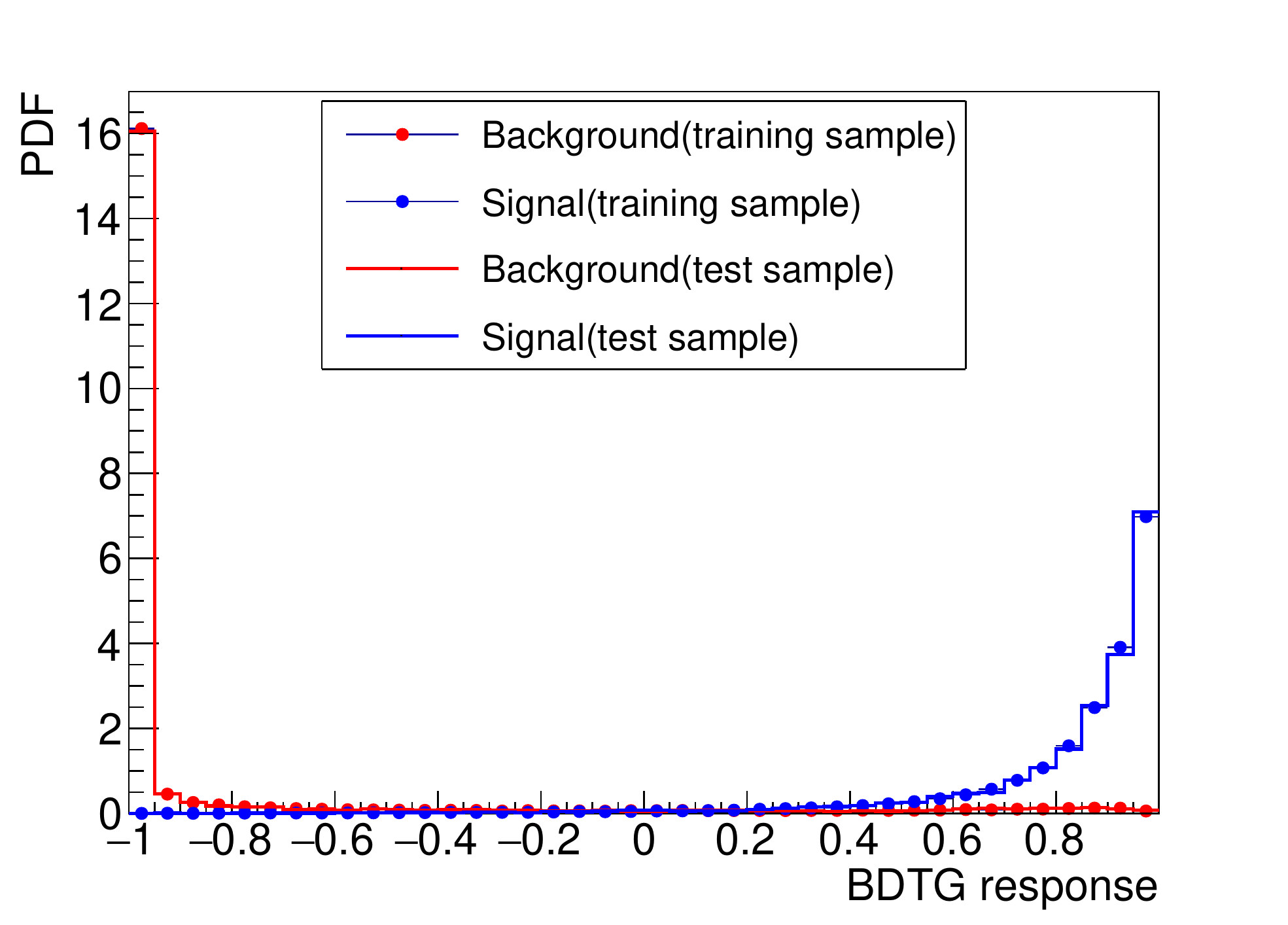}
            \label{fig:response_BDTG}
        }
        \quad
        \subfigure[]{
            \includegraphics[width=0.4\hsize]{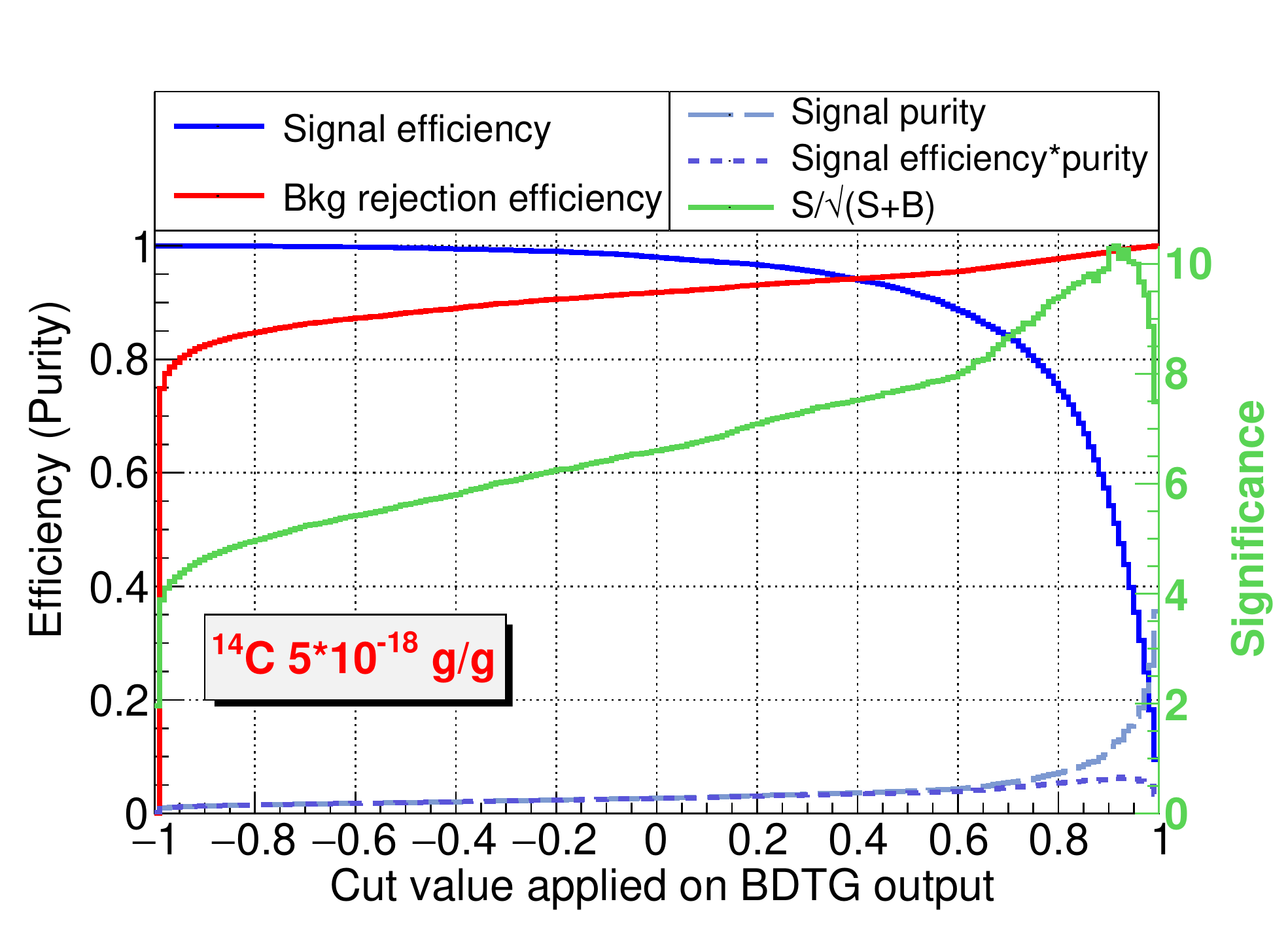}
            \label{fig:efficiency_BDTG_5e-18}
        }
        \quad
        \subfigure[]{
            \includegraphics[width=0.4\hsize]{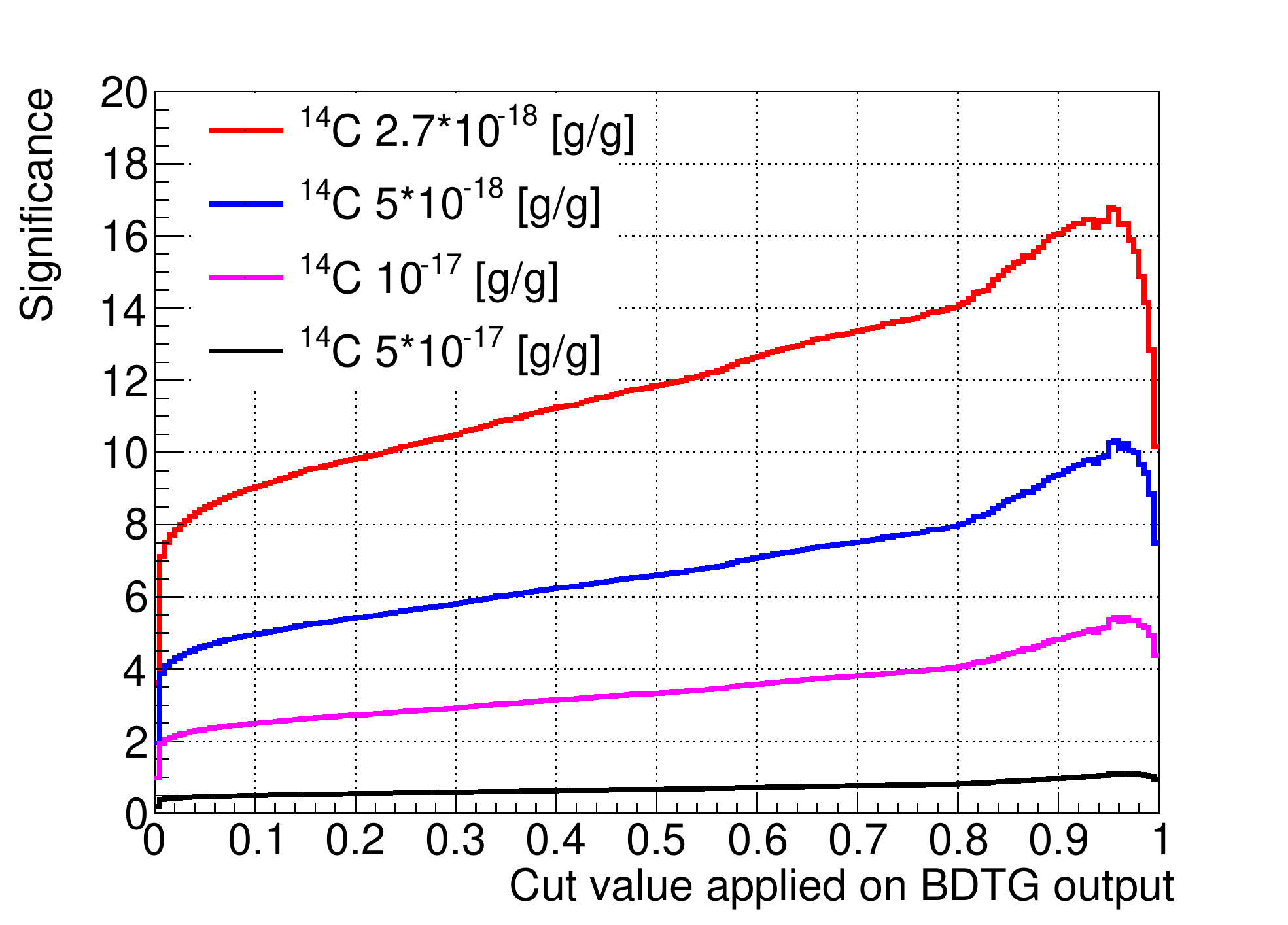}
            \label{fig:significance_BDTG_differentC14}
        }
        \quad
        \subfigure[]{
            \includegraphics[width=0.4\hsize]{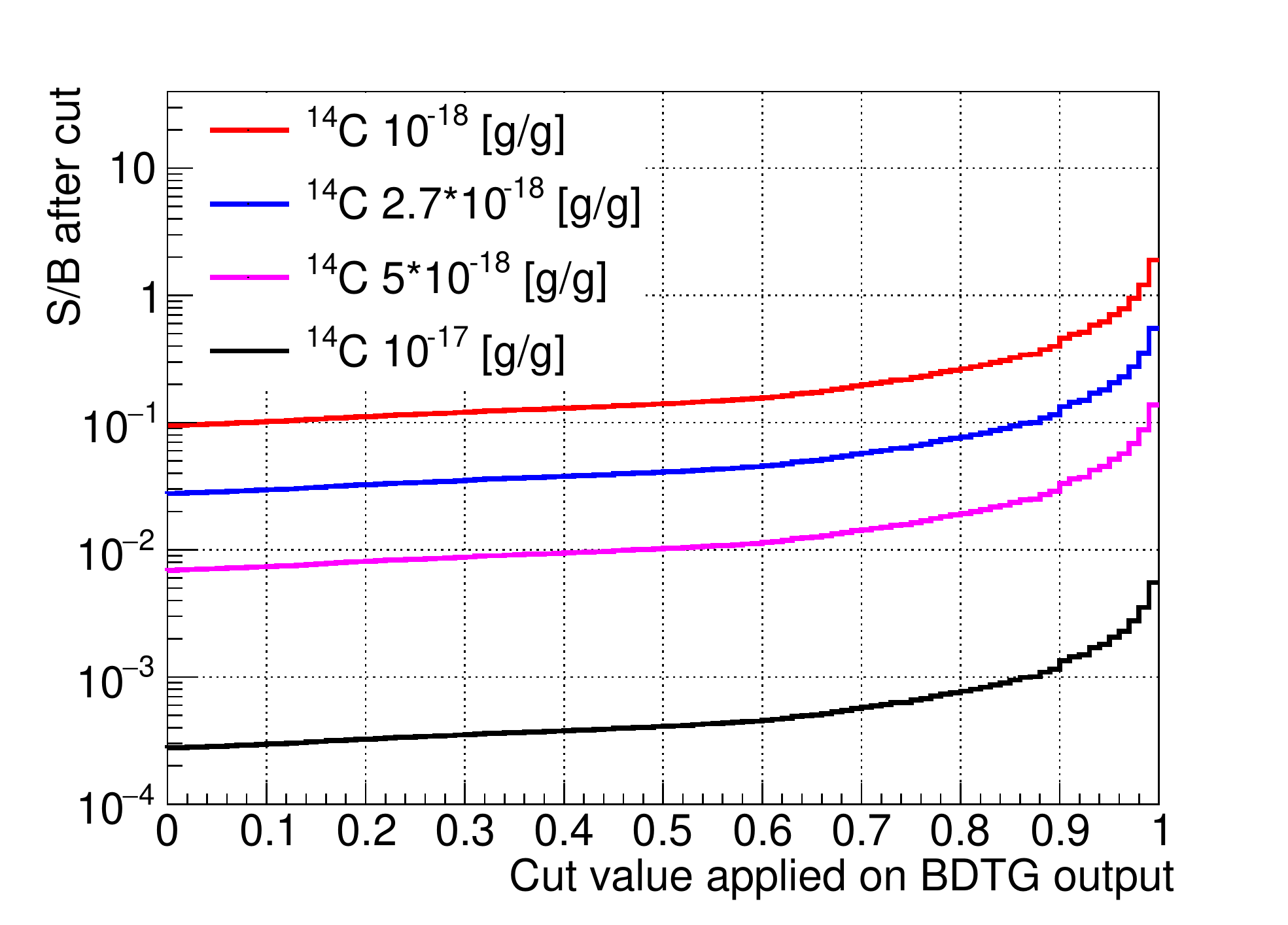}
            \label{fig:StoB_afterCut_BDTG_differentC14}
        }
        \caption{Identification performance using the BDTG model. (a) Normalized response distributions of the BDTG model for the signal and the background. (b) Cut efficiencies as functions of BDTG cut values. The significance (green line) was calculated using one day of statistics of the signal and the background in the analysis region, and the $^{14}$C concentration of LS is assumed to $5 \times 10^{-18}$~g/g. (c) Significance for different assumptions of $^{14}$C concentration. (d) Signal-to-background ratio after identification in the case of different assumptions of $^{14}$C concentration, one day of statistics were adopted.}
        \label{fig:BDTG_distribution}
    \end{figure*}
    
    Fig.~\ref{fig:BDTG_distribution} shows the training results of the BDTG model. The network is not overtrained as the response of testing data is consistent with the training data (Fig.~\ref{fig:response_BDTG}. Basically, the signal and the background are separated into two parts after training, but there are still some overlapping components, indicating that their event features are similar and hence the network fails to distinguish between them. According to a detailed investigation, it was found that one of the main contributions to the failed identification comes from the stacking case of two $^{14}$C that are very close together in both time and space. To optimize the significance: $N_s/\sqrt{N_s+N_b}$ (where $N_s$ and $N_b$ are the numbers of signal and background after identification), we scanned the cut value on BDTG response and the corresponding efficiencies can be obtained as well. The $^{14}$C concentration of LS is assumed to $5 \times 10^{-18}$~g/g in Fig.~\ref{fig:efficiency_BDTG_5e-18}, the calculation of the significance using one day of statistics in the analysis region (true energy: 160-250~keV) based on the estimation in Table~\ref{table:tab-eventRate}, they are $\sim$1653 for signal and $\sim$712440 for background (only consider $^{14}$C double pile-up events) before the identification. For the BDTG model, the significance can reach its maximum value of 10.33 after applying a cut at 0.915, and the signal efficiency and the background rejection efficiency are 51.1\% and 99.18\% in this case. As discussed in Sec.~\ref{section:Introduction}, the signal-to-background ratio of $pp$ neutrinos and $^{14}$C double pile-up events is poor in a large-scale LS detector, thus a strict cut is needed to reject the most of background. In this case, 51.1\% is an acceptable value for signal efficiency, and it still corresponds to a much larger statistics of effective \textit{pp} neutrino signal per day compared to most existing experiments.
    
	In Fig.~\ref{fig:significance_BDTG_differentC14}, significance is evaluated using different assumptions of $^{14}$C concentration, while Fig.~\ref{fig:StoB_afterCut_BDTG_differentC14} shows the signal-to-background ratio after identification using the BDTG model, and the calculations were based on one day of statistics in the case of different $^{14}$C concentrations. As a result, the BDTG model shows great performance and it can handle most of $^{14}$C double pile-up events effectively.
	
	\begin{figure}[!htb]
        \flushleft
		\centering
		\includegraphics[width=0.9\hsize]{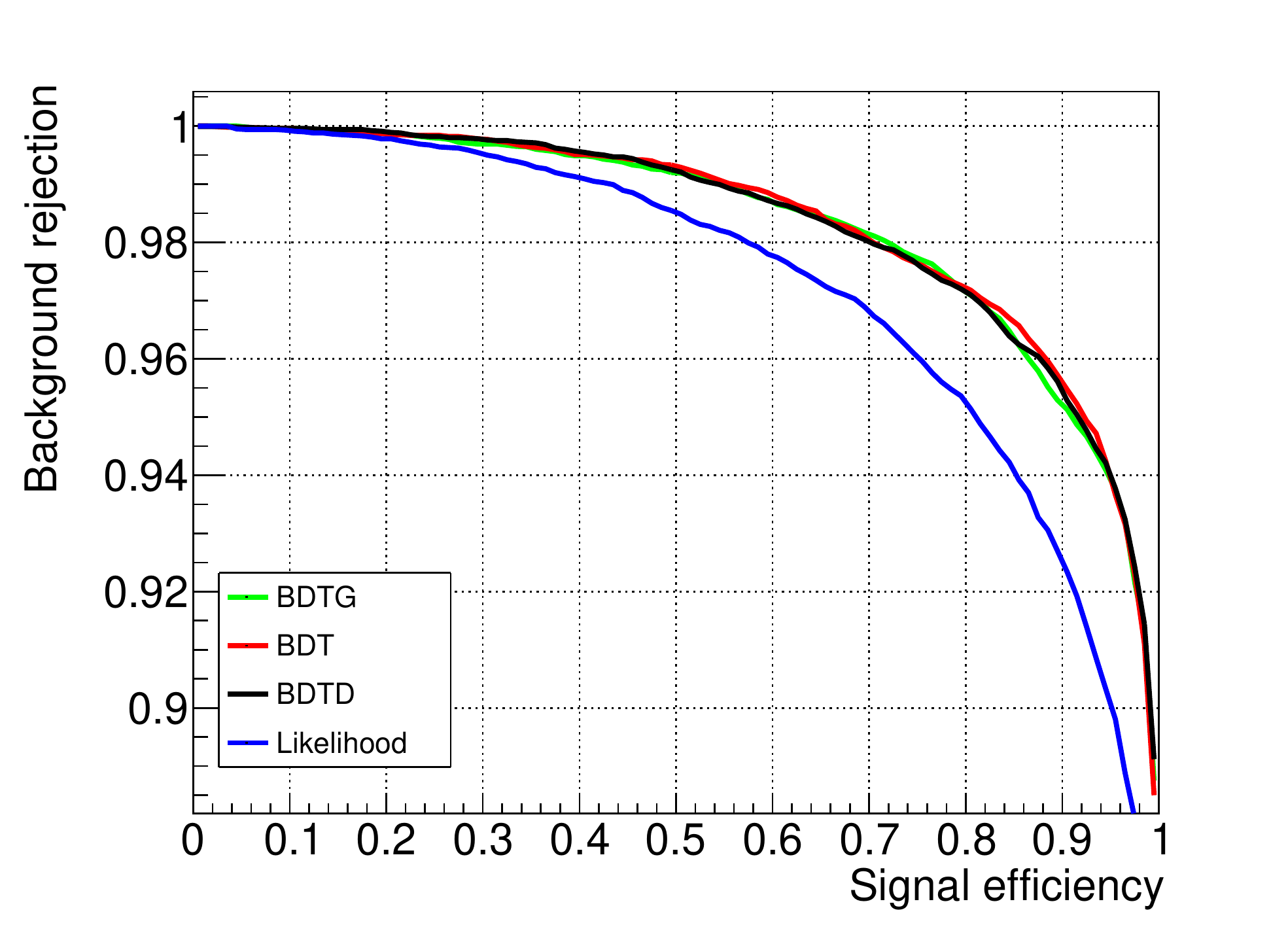}
		\caption{Relation of background rejection efficiency to signal efficiency for several TMVA algorithms.} 
		\label{fig:TMVA_compare} 
	\end{figure}
	
	\begin{figure*}[!htb]
        \centering
        \subfigure[]{
            \includegraphics[width=0.4\hsize]{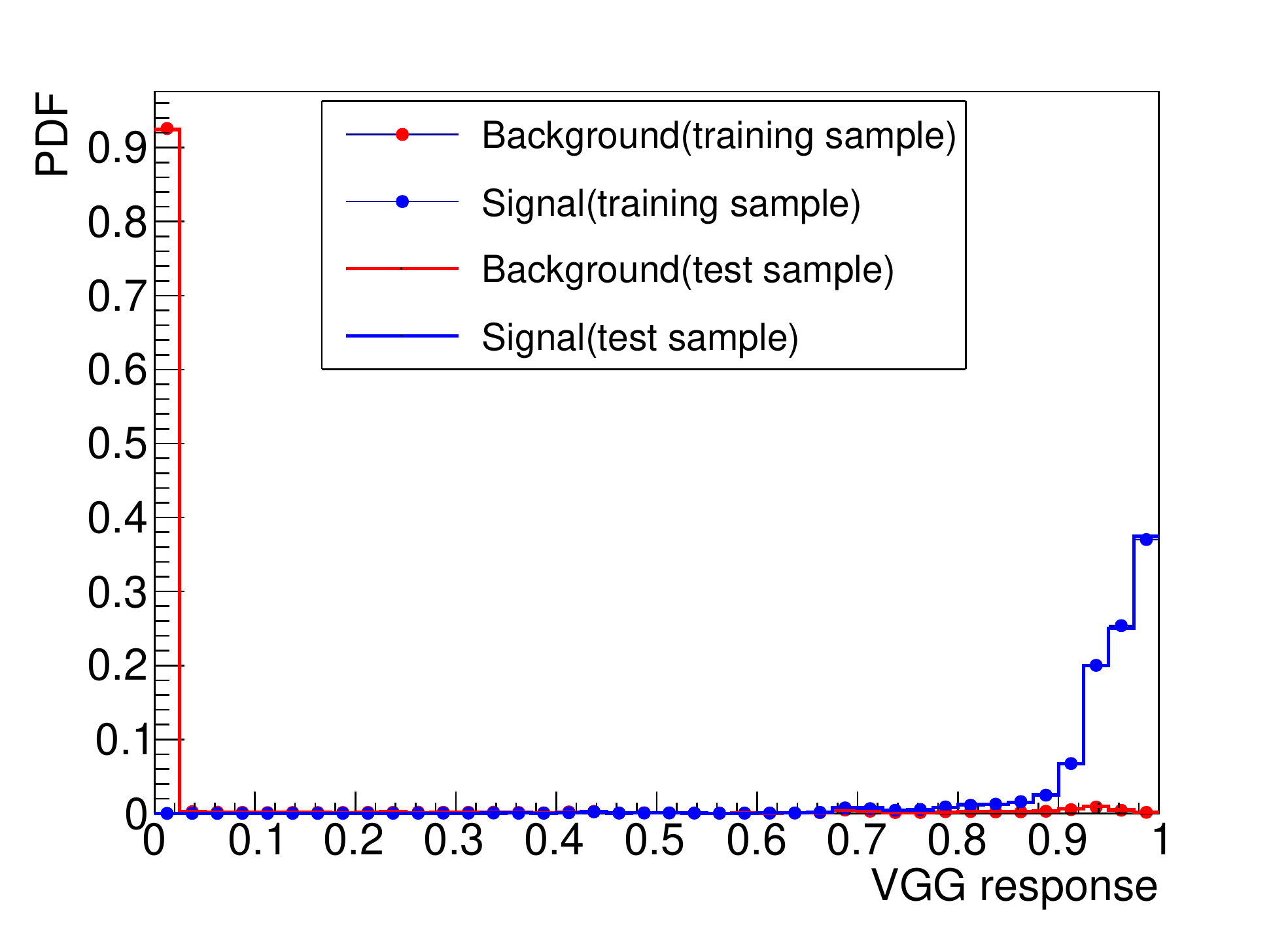}
            \label{fig:response_VGG}
        }
        \quad
        \subfigure[]{
            \includegraphics[width=0.4\hsize]{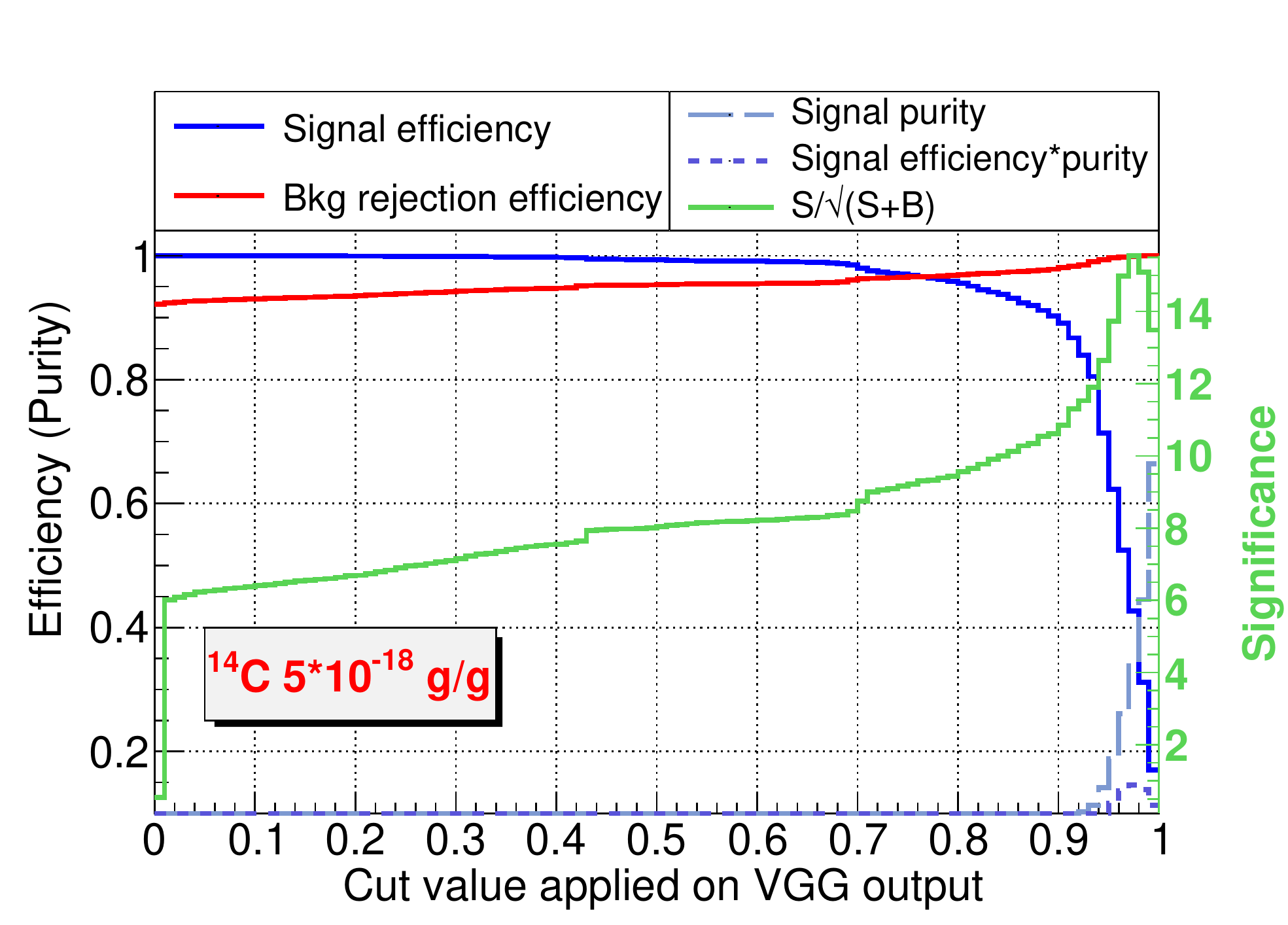}
            \label{fig:efficiency_VGG_5e-18}
        }
        \quad
        \subfigure[]{
            \includegraphics[width=0.4\hsize]{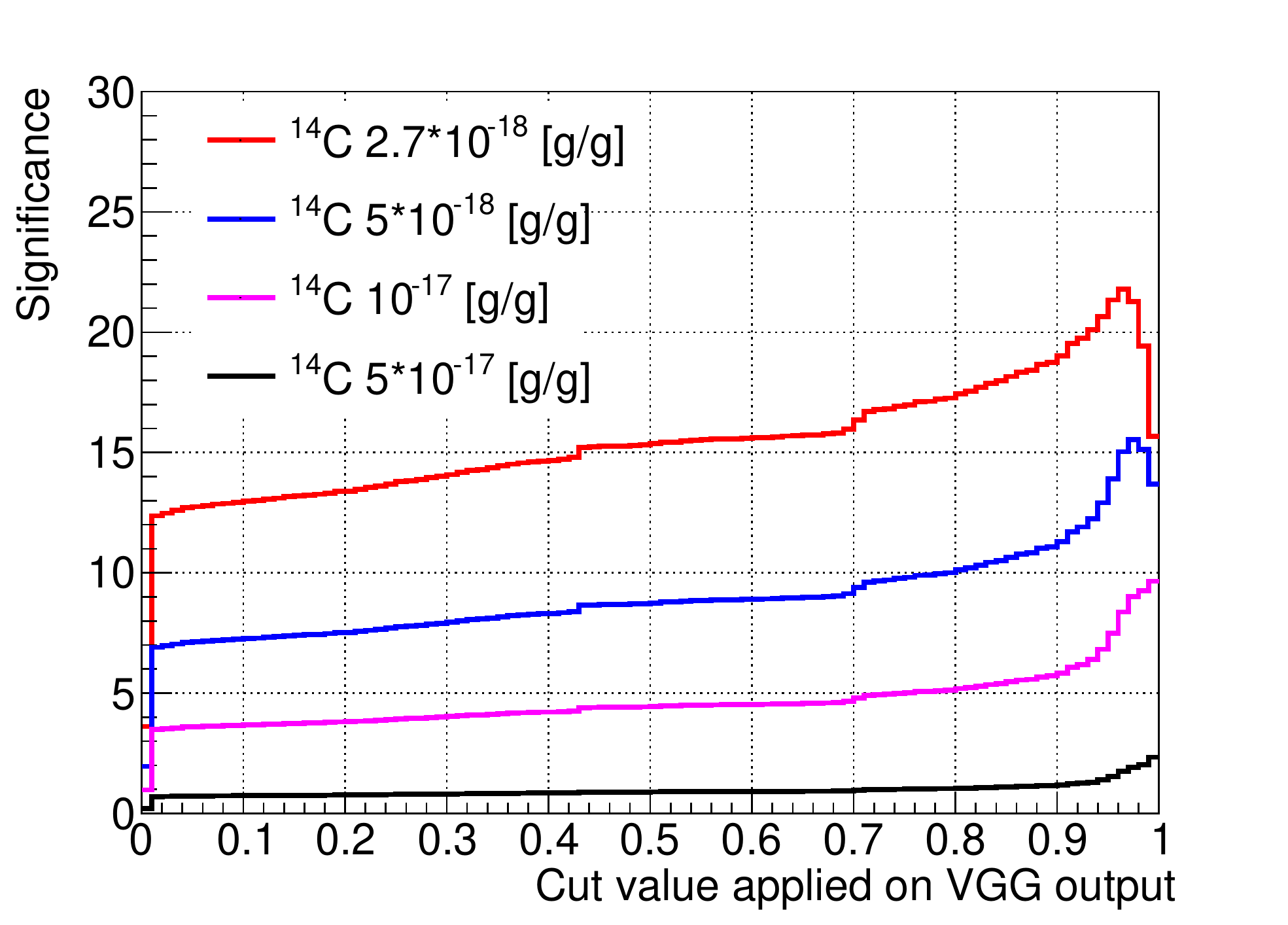}
            \label{fig:significance_VGG_differentC14}
        }
        \quad
        \subfigure[]{
            \includegraphics[width=0.4\hsize]{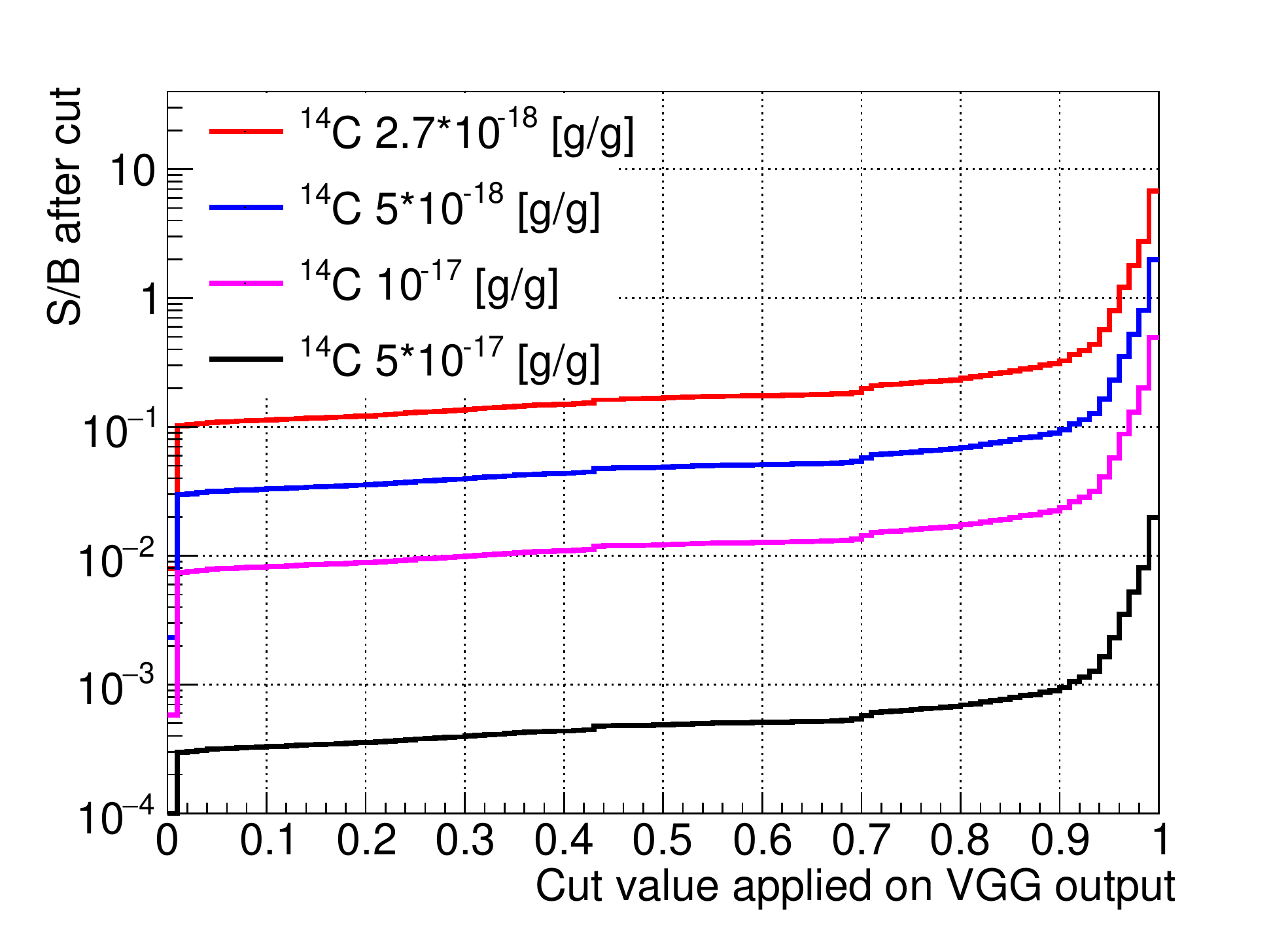}
            \label{fig:StoB_afterCut_VGG_differentC14}
        }
        \caption{Identification performance using the VGG network. (a) Normalized response distributions of the VGG network for the signal and the background. (b) Cut efficiencies as functions of VGG cut values. The significance (green line) was calculated using one day of statistics of the signal and the background in the analysis region, and the $^{14}$C concentration of LS is assumed to $5 \times 10^{-18}$~g/g. (c) Significance for different assumptions of $^{14}$C concentration. (d) Signal-to-background ratio after identification in the case of different assumptions of $^{14}$C concentration, one day of statistics were adopted.}
        \label{fig:VGG_distribution}
    \end{figure*}

    In addition, other TMVA algorithms are also investigated, including the Likelihood algorithm and several BDT models (BDT, BDTD). Many of them exhibit similar performance (Fig.~\ref{fig:TMVA_compare}), indicating the good robustness and stability of our analysis. 

\subsection{Discrimination performance of the VGG network}
\label{subsection:result_VGG}

    Fig.~\ref{fig:VGG_distribution} shows the training results of the VGG network. The network is not overtrained as the response of testing data is consistent with the training data (Fig.~\ref{fig:response_VGG}. To optimize the significance, we scan the cut values on the VGG output, and the corresponding efficiencies can also be obtained. The $^{14}$C concentration of LS is assumed to $5 \times 10^{-18}$~g/g in Fig.~\ref{fig:efficiency_VGG_5e-18}, and the calculation of the significance using one day of statistics in the analysis region based on the estimation in Table~\ref{table:tab-eventRate}. For the VGG network, the significance can reach its maximum value of 15.55 after applying a cut at 0.975, and the signal efficiency and the background rejection efficiency are 42.7\% and 99.81\% in this case. 
    
	In Fig.~\ref{fig:significance_VGG_differentC14}, significance is evaluated using different assumptions of $^{14}$C concentration, while Fig.~\ref{fig:StoB_afterCut_VGG_differentC14} shows the signal-to-background ratio after identification using the VGG network, and the calculations were based on one day of statistics in the case of different $^{14}$C concentrations. As a result, the VGG network shows great performance and it is able to achieve higher significance and a good improvement in the signal-to-background ratio compared to the BDTG model.
	
	\begin{figure}[!htb]
        \flushleft
		\centering
		\includegraphics[width=0.9\hsize]{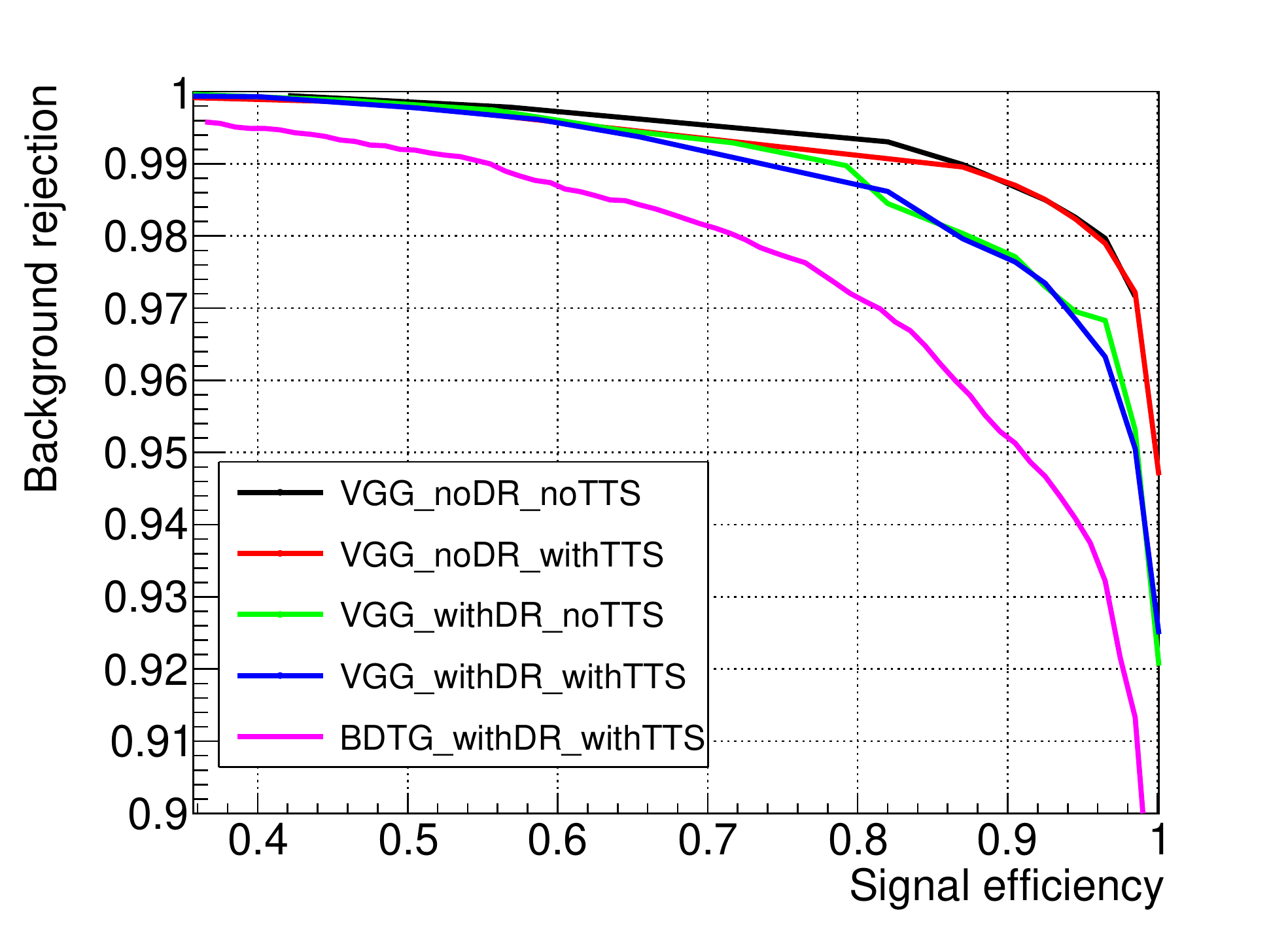}
		\caption{Relation of background rejection efficiency to signal efficiency for different MC samples.} 
		\label{fig:roc_differentCase} 
	\end{figure}
	
	Furthermore, the discrimination performance was compared using different MC samples, as shown in Fig.~\ref{fig:roc_differentCase}, the performance of discrimination gets worse after included the PMT dark noise, while TTS only has a little influence. And the discrimination performance based on the VGG network is stable when rejecting $\sim$99.8\% $^{14}$C double pile-up events.

\section{Summary}
\label{section:Summary}
	
    Large-scale LS detectors have the benefits of huge target mass and high energy resolution, which makes them have a good potential in \textit{pp} solar neutrino detection, but also face serious $^{14}$C pile-up background. In this paper, we investigate the discrimination of \textit{pp} solar neutrinos and $^{14}$C double pile-up events in a large-scale LS detector using both multivariate analysis and deep learning technology. In the simulation study, a spherical LS detector was built using the Geant4 toolkit, and comprehensive optical processes were adopted. The response features in the PMT hit patterns of \textit{pp} neutrinos and $^{14}$C double pile-up events were compared, clear differences were found in their time and spatial distributions since one of them is a single point-like event and the other one is an accidental coincidence of multiple events.
    
    For the discrimination based on the BDTG model, a signal significance of 10.3 can be achieved using only one day of statistics. And the signal efficiency is 51.1\% when rejecting 99.18\% $^{14}$C double pile-up events. As for the VGG network, signal significance can achieve 15.6 using only one day of statistics, and the signal efficiency is 42.7\% when rejecting 99.81\% $^{14}$C double pile-up events. This analysis provides a reliable method reference for similar experiments in low-threshold physics detection and $^{14}$C pile-up background reduction.


\begin{thebibliography}{99}
\bibitem{Bethe-1938} 
H.A.~Bethe and C.L.~Critchfield, The formation of deuterium by proton combination. Phys. Rev. \textbf{54}, 248-254 (1938). \href{https://doi.org/10.1103/PhysRev.54.248}{https://doi.org/10.1103/PhysRev.54.248}

\bibitem{Bethe-1939} 
H.A.~Bethe, Energy production in stars. Phys. Rev. 55, 434–456 (1939). \href{https://doi.org/10.1103/PhysRev.55.434}{https://doi.org/10.1103/PhysRev.55.434}

\bibitem{Bahcall-1996pt}
J.N.~Bahcall, M.~Fukugita and P.I.~Krastev, How does the Sun shine? Phys. Lett. B \textbf{374}, 1-6 (1996). \href{https://doi.org/10.1016/0370-2693(96)00187-6}{https://doi.org/10.1016/0370-2693(96)00187-6}

\bibitem{Davis-1968cp}
R.~Davis, Jr., D.S.~Harmer \textit{et al.}, Search for neutrinos from the sun. Phys. Rev. Lett. \textbf{20}, 1205-1209 (1968). \href{https://doi.org/10.1103/PhysRevLett.20.1205}{https://doi.org/10.1103/PhysRevLett.20.1205}

\bibitem{Cleveland-1998nv}
B.T.~Cleveland, T.~Daily, R.~Davis \textit{et al.}, Measurement of the solar electron neutrino flux with the Homestake chlorine detector. Astrophys. J. \textbf{496}, 505-526 (1998). \href{https://doi.org/10.1086/305343}{https://doi.org/10.1086/305343}

\bibitem{GALLEX-1992gcp}
P.~Anselmann, W.~Hampel, G.~Heusser \textit{et al.}, Solar neutrinos observed by GALLEX at Gran Sasso. Phys. Lett. B \textbf{285}, 376-389 (1992). \href{https://doi.org/10.1016/0370-2693(92)91521-A}{https://doi.org/10.1016/0370-2693(92)91521-A}

\bibitem{GALLEX-1998kcz}
W.~Hampel, J.~Handt, G.~Heusser \textit{et al.}, GALLEX solar neutrino observations: Results for GALLEX IV. Phys. Lett. B \textbf{447}, 127-133 (1999). \href{https://doi.org/10.1016/S0370-2693(98)01579-2}{https://doi.org/10.1016/S0370-2693(98)01579-2}

\bibitem{Kaether-2010ag}
F.~Kaether, W.~Hampel, G.~Heusser \textit{et al.}, Reanalysis of the GALLEX solar neutrino flux and source experiments. Phys. Lett. B \textbf{685}, 47-54 (2010). \href{https://doi.org/10.1016/j.physletb.2010.01.030}{https://doi.org/10.1016/j.physletb.2010.01.030}

\bibitem{GNO-2005bds}
M.~Altmann, M.~Balata, P.~Belli \textit{et al.}, Complete results for five years of GNO solar neutrino observations. Phys. Lett. B \textbf{616}, 174-190 (2005). \href{https://doi.org/10.1016/j.physletb.2005.04.068}{https://doi.org/10.1016/j.physletb.2005.04.068}

\bibitem{SAGE-2009eeu}
J.N.~Abdurashitov, V.N.~Gavrin, V.V.~Gorbachev \textit{et al.}, Measurement of the solar neutrino capture rate with gallium metal. III: Results for the 2002--2007 data-taking period. Phys. Rev. C \textbf{80}, 015807 (2009). \href{https://doi.org/10.1103/PhysRevC.80.015807}{https://doi.org/10.1103/PhysRevC.80.015807}

\bibitem{Gavrin-2019sok}
V.~N.~Gavrin, The history, present and future of SAGE (Soviet-American Gallium Experiment). \href{https://doi.org/10.1142/9789811204296\_0002}{https://doi.org/10.1142/9789811204296\_0002}

\bibitem{Kamiokande-II-1989hkh}
K.S.~Hirata, T.~Kajita, T. Kifune \textit{et al.}, Observation of B-8 Solar Neutrinos in the Kamiokande-II Detector. Phys. Rev. Lett. \textbf{63}, 16 (1989). \href{https://doi.org/10.1103/PhysRevLett.63.16}{https://doi.org/10.1103/PhysRevLett.63.16}

\bibitem{Kamiokande-1996qmi}
Y.~Fukuda, T.~Hayakawa, K. Inoue \textit{et al.}, Solar neutrino data covering solar cycle 22, Phys. Rev. Lett. \textbf{77}, 1683-1686 (1996). \href{https://doi.org/10.1103/PhysRevLett.77.1683}{https://doi.org/10.1103/PhysRevLett.77.1683}

\bibitem{Wolfenstein-1977ue}
L.~Wolfenstein, Neutrino Oscillations in Matter. Phys. Rev. D \textbf{17}, 2369-2374 (1978). \href{https://doi.org/10.1103/PhysRevD.17.2369}{https://doi.org/10.1103/PhysRevD.17.2369}

\bibitem{Mikheyev-1985zog}
S.P.~Mikheyev and A.Y.~Smirnov, Resonance Amplification of Oscillations in Matter and Spectroscopy of Solar Neutrinos. Sov. J. Nucl. Phys. \textbf{42}, 913-917 (1985). 

\bibitem{SNO-2001kpb}
Q.R.~Ahmad, R.C.~Allen, J.D.~Anglin \textit{et al.}, Measurement of the rate of $\nu_e+d \to p+p+e^-$ interactions produced by $^8$B solar neutrinos at the Sudbury Neutrino Observatory. Phys. Rev. Lett. \textbf{87}, 071301 (2001). \href{https://doi.org/10.1103/PhysRevLett.87.071301}{https://doi.org/10.1103/PhysRevLett.87.071301}

\bibitem{SNO-2003bmh}
S.N.~Ahmed, A.E.~Anthony, E.W.~Beier \textit{et al.}, Measurement of the total active B-8 solar neutrino flux at the Sudbury Neutrino Observatory with enhanced neutral current sensitivity. Phys. Rev. Lett. \textbf{92}, 181301 (2004). \href{https://doi.org/10.1103/PhysRevLett.92.181301}{https://doi.org/10.1103/PhysRevLett.92.181301}

\bibitem{KamLAND-2002uet}
K.~Eguchi, S.~Enomoto, K.~Furuno \textit{et al.}, First results from KamLAND: Evidence for reactor anti-neutrino disappearance. Phys. Rev. Lett. \textbf{90}, 021802 (2003). \href{https://doi.org/10.1103/PhysRevLett.90.021802}{https://doi.org/10.1103/PhysRevLett.90.021802}

\bibitem{Bahcall-1995bt}
J.N.~Bahcall and M.H.~Pinsonneault, Solar models with helium and heavy element diffusion. Rev. Mod. Phys. \textbf{67}, 781-808 (1995). \href{https://doi.org/10.1103/RevModPhys.67.781}{https://doi.org/10.1103/RevModPhys.67.781}

\bibitem{Christensen-Dalsgaard-1996hpz}
J.~Christensen-Dalsgaard, W.~Dappen, S.V.~Ajukov \textit{et al.}, The current state of solar modeling. Science \textbf{272}, 1286-1292 (1996). \href{https://doi.org/10.1126/science.272.5266.1286}{https://doi.org/10.1126/science.272.5266.1286}

\bibitem{DeglInnocenti-1996uex}
S.~Degl'Innocenti, W.A.~Dziembowski, G.~Fiorentini \textit{et al.}, Helioseismology and standard solar models. Astropart. Phys. \textbf{7}, 77-95 (1997). \href{https://doi.org/10.1016/S0927-6505(97)00004-2}{https://doi.org/10.1016/S0927-6505(97)00004-2}

\bibitem{Brun-1999dw}
A.S.~Brun, S.~Turck-Chieze and J.~P.~Zahn, Standard solar models in the light of new helioseismic constraints. 2. mixing below the convective zone. Astrophys. J. \textbf{525}, 1032-1041 (1999). \href{https://doi.org/10.1086/307932}{https://doi.org/10.1086/307932}

\bibitem{Bahcall-2001pf}
J.N.~Bahcall, The Luminosity constraint on solar neutrino fluxes. Phys. Rev. C \textbf{65}, 025801 (2002). \href{https://doi.org/10.1103/PhysRevC.65.025801}{https://doi.org/10.1103/PhysRevC.65.025801}

\bibitem{Serenelli-2009yc}
A.~Serenelli, S.~Basu, J.W.~Ferguson \textit{et al.}, New Solar Composition: The Problem With Solar Models Revisited. Astrophys. J. Lett. \textbf{705}, L123-L127 (2009). \href{https://doi.org/10.1088/0004-637X/705/2/L123}{https://doi.org/10.1088/0004-637X/705/2/L123}

\bibitem{Borexino-2007kvk}
C.~Arpesella, G.~Bellini, J.~Benziger \textit{et al.}, First real time detection of Be-7 solar neutrinos by Borexino. Phys. Lett. B \textbf{658}, 101-108 (2008). \href{https://doi.org/10.1016/j.physletb.2007.09.054}{https://doi.org/10.1016/j.physletb.2007.09.054}

\bibitem{BOREXINO-2014pcl}
G.~Bellini, J.~Benziger, D.~Bick \textit{et al.}, Neutrinos from the primary proton\textendash{}proton fusion process in the Sun. Nature \textbf{512}, no.7515, 383-386 (2014). \href{https://doi.org/10.1038/nature13702}\href{https://doi.org/10.1038/nature13702}

\bibitem{BOREXINO-2018ohr}
M.~Agostini, K.~Altenmüller, S.~Appel \textit{et al.}, Comprehensive measurement of $pp$-chain solar neutrinos. Nature \textbf{562}, no.7728, 505-510 (2018). \href{https://doi.org/10.1038/s41586-018-0624-y}{https://doi.org/10.1038/s41586-018-0624-y}

\bibitem{BOREXINO-2020aww}
M.~Agostini, K.~Altenmüller, S.~Appel \textit{et al.}, Experimental evidence of neutrinos produced in the CNO fusion cycle in the Sun. Nature \textbf{587}, 577-582 (2020). \href{https://doi.org/10.1038/s41586-020-2934-0}{https://doi.org/10.1038/s41586-020-2934-0}



\bibitem{Gann-2021ndb}
G.D.O.~Gann, K.~Zuber, D.~Bemmerer \textit{et al.}, The Future of Solar Neutrinos. Ann. Rev. Nucl. Part. Sci. \textbf{71}, 491-528 (2021). \href{https://doi.org/10.1146/annurev-nucl-011921-061243}{https://doi.org/10.1146/annurev-nucl-011921-061243}

\bibitem{Xu-2022wcq}
X.J.~Xu, Z.~Wang and S.~Chen, Solar neutrino physics. \href{
https://doi.org/10.48550/arXiv.2209.14832}{https://doi.org/10.48550/arXiv.2209.14832}

\bibitem{JUNO-PPNP}
A. Abusleme \textit{et al.}, JUNO Physics and Detector. Prog. Part. Nucl. Phys. \textbf{123}, 103927 (2022). \href{https://doi.org/10.1016/j.ppnp.2021.103927}{https://doi.org/10.1016/j.ppnp.2021.103927}

\bibitem{Jinping-2016iiq}
J.F.~Beacom, S.M.~Chen, J.P.~Cheng \textit{et al.}, Physics prospects of the Jinping neutrino experiment. Chin. Phys. C \textbf{41}, no.2, 023002 (2017). \href{https://doi.org/10.1088/1674-1137/41/2/023002}{https://doi.org/10.1088/1674-1137/41/2/023002}

\bibitem{DARWIN-2020bnc}
J.~Aalbers, F.~Agostini, S.E.M.~Ahmed Maouloud \textit{et al.}, Solar neutrino detection sensitivity in DARWIN via electron scattering. Eur. Phys. J. C \textbf{80}, no.12, 1133 (2020). \href{https://doi.org/10.1140/epjc/s10052-020-08602-7}{https://doi.org/10.1140/epjc/s10052-020-08602-7}

\bibitem{Bieger-2021sas}
L.~Bieger, T.~Birkenfeld, D.~Blum \textit{et al.}, Potential for a precision measurement of solar pp neutrinos in the Serappis experiment. Eur. Phys. J. C \textbf{82}, no.9, 779 (2022). \href{https://doi.org/10.1140/epjc/s10052-022-10725-y}{https://doi.org/10.1140/epjc/s10052-022-10725-y}

\bibitem{juno-yellowbook}
F.P. An, G.P An, Q. An \textit{et al.}, Neutrino Physics with JUNO. J. Phys. G \textbf{43}, no.3, 030401 (2016). \href{https://doi.org/10.1088/0954-3899/43/3/030401}{https://doi.org/10.1088/0954-3899/43/3/030401}

\bibitem{LENA}
M. Wurm, Studying neutrino properties in the future LENA experiment. Nucl. Phys. B Proc. Suppl. \textbf{237-238}, 314-316 (2013). \href{https://doi.org/10.1016/j.nuclphysbps.2013.04.114}{https://doi.org/10.1016/j.nuclphysbps.2013.04.114}

\bibitem{GEANT4-2002zbu}
S.~Agostinelli, J.~Allison, K.~Amako \textit{et al.}, GEANT4--a simulation toolkit. Nucl. Instrum. Meth. A \textbf{506}, 250-303 (2003). \href{https://doi.org/10.1016/S0168-9002(03)01368-8}{https://doi.org/10.1016/S0168-9002(03)01368-8}

\bibitem{Zhou-2015gwa}
X.~Zhou, Q.~Liu, M.~Wurm \textit{et al.}, Rayleigh scattering of linear alkylbenzene in large liquid scintillator detectors. Rev. Sci. Instrum. \textbf{86}, no.7, 073310 (2015). \href{https://doi.org/10.1063/1.4927458}{https://doi.org/10.1063/1.4927458}

\bibitem{Gao-2013pua}
L.~Gao, B.x.~Yu, Y.y.~Ding \textit{et al.}, Attenuation length measurements of a liquid scintillator with LabVIEW and reliability evaluation of the device. Chin. Phys. C \textbf{37}, 076001 (2013). \href{https://doi.org/10.1088/1674-1137/37/7/076001}{https://doi.org/10.1088/1674-1137/37/7/076001}

\bibitem{Wurm-2010ad}
M.~Wurm, F.von Feilitzsch, M.~Goeger-Neff \textit{et al.}, Optical Scattering Lengths in Large Liquid-Scintillator Neutrino Detectors. Rev. Sci. Instrum. \textbf{81}, 053301 (2010). \href{https://doi.org/10.1063/1.3397322}{https://doi.org/10.1063/1.3397322}

\bibitem{Zhang-2020mqz}
Y.~Zhang, Z.Y.~Yu, X.Y.~Li \textit{et al.}, A complete optical model for liquid-scintillator detectors. Nucl. Instrum. Meth. A \textbf{967}, 163860 (2020). \href{https://doi.org/10.1016/j.nima.2020.163860}{https://doi.org/10.1016/j.nima.2020.163860}

\bibitem{Ding-2015sys}
X.F.~Ding, L.J.~Wen, X.~Zhou \textit{et al.}, Measurement of the fluorescence quantum yield of bis-MSB. Chin. Phys. C \textbf{39}, no.12, 126001 (2015). \href{https://doi.org/10.1088/1674-1137/39/12/126001}{https://doi.org/10.1088/1674-1137/39/12/126001}

\bibitem{Buck-2015jxa}
C.~Buck, B.~Gramlich and S.~Wagner, Light propagation and fluorescence quantum yields in liquid scintillators. JINST \textbf{10}, no.09, P09007 (2015). \href{https://doi.org/10.1088/1748-0221/10/09/P09007}{https://doi.org/10.1088/1748-0221/10/09/P09007}

\bibitem{OKeeffe-2011dex}
H.M.~O'Keeffe, E.~O'Sullivan and M.C.~Chen, Scintillation decay time and pulse shape discrimination in oxygenated and deoxygenated solutions of linear alkylbenzene for the SNO+ experiment. Nucl. Instrum. Meth. A \textbf{640}, 119-122 (2011). \href{https://doi.org/10.1016/j.nima.2011.03.027}{https://doi.org/10.1016/j.nima.2011.03.027}

\bibitem{Yu-2022god}
M.~Yu, L.~Wen, X.~Zhou \textit{et al.}, Determine Energy Nonlinearity and Resolution of $e^{\pm}$ and $\gamma$ in Liquid Scintillator Detectors by A Universal Energy Response Model. \href{https://doi.org/10.48550/arXiv.2211.02467}{https://doi.org/10.48550/arXiv.2211.02467}

\bibitem{DayaBay-2019fje}
D.~Adey, A.B.~Balantekin, M.~Bishai \textit{et al.}, A high precision calibration of the nonlinear energy response at Daya Bay. Nucl. Instrum. Meth. A \textbf{940}, 230-242 (2019). \href{https://doi.org/10.1016/j.nima.2019.06.031}{https://doi.org/10.1016/j.nima.2019.06.031}

\bibitem{Dunger:2019dfo}
J. Dunger and S.D. Biller, Multi-site Event Discrimination in Large Liquid Scintillation Detectors, Nucl. Instrum. Meth. A \textbf{943}, 162420 (2019). \href{https://doi.org/10.1016/j.nima.2019.162420}{https://doi.org/10.1016/j.nima.2019.162420}

\bibitem{Hocker:2007ht}
A. Hocker, P. Speckmayer, J. Stelzer \textit{et al.}, TMVA - Toolkit for Multivariate Data Analysis, \href{https://doi.org/10.48550/arXiv.physics/0703039}{https://doi.org/10.48550/arXiv.physics/0703039} 

\bibitem{Speckmayer:2010zz}
P. Speckmayer, A. Hocker, J. Stelzer \textit{et al.}, The toolkit for multivariate data analysis, TMVA 4, J. Phys. Conf. Ser. \textbf{219}, 032057 (2010). \href{https://doi.org/10.1088/1742-6596/219/3/032057}{https://doi.org/10.1088/1742-6596/219/3/032057}

\bibitem{Lampen:2008zza}
T. Lampen, F. Garcia, A. Heikkinen \textit{et al.}, Testing TMVA software in b-tagging for the search of MSSM Higgs bosons at the LHC, J. Phys. Conf. Ser. \textbf{119}, 032028 (2008). \href{https://doi.org/10.1088/1742-6596/119/3/032028}{https://doi.org/10.1088/1742-6596/119/3/032028}

\bibitem{LHAASO:2019qdu}
L.Q.~Yin, S.S.~Zhang, Z.~Cao \textit{et al.} [LHAASO], Expected energy spectrum of cosmic ray protons and helium below 4 PeV measured by LHAASO, Chin. Phys. C \textbf{43}, no.7, 075001 (2019). \href{https://doi.org/10.1088/1674-1137/43/7/075001}{https://doi.org/10.1088/1674-1137/43/7/075001}

\bibitem{Guest-2016iqz}
D.~Guest, J.~Collado, P.~Baldi \textit{et al.}, Jet Flavor Classification in High-Energy Physics with Deep Neural Networks, Phys. Rev. D \textbf{94}, no.11, 112002 (2016). \href{https://doi.org/10.1103/PhysRevD.94.112002}{https://doi.org/10.1103/PhysRevD.94.112002}

\bibitem{Guest-2018yhq}
D.~Guest, K.~Cranmer and D.~Whiteson, Deep Learning and its Application to LHC Physics, Ann. Rev. Nucl. Part. Sci. \textbf{68}, 161-181 (2018). \href{https://doi.org/10.1146/annurev-nucl-101917-021019}{https://doi.org/10.1146/annurev-nucl-101917-021019}

\bibitem{He-2018nst}
J.P.~He, X.B.~Tang, P.~Gong \textit{et al.}, Spectrometry analysis based on approximation coefficients and deep belief networks. NUCL SCI TECH 29, 69 (2018). \href{https://doi.org/10.1007/s41365-018-0402-4}{https://doi.org/10.1007/s41365-018-0402-4}

\bibitem{Ma-2019nst}
X.K.~Ma, H.Q.~Huang, Q.C.~Wang \textit{et al.}, Estimation of Gaussian overlapping nuclear pulse parameters based on a deep learning LSTM model. NUCL SCI TECH 30, 171 (2019). \href{https://doi.org/10.1007/s41365-019-0691-2}{https://doi.org/10.1007/s41365-019-0691-2}

\bibitem{Qian-2021vnh}
Z.~Qian, V.~Belavin, V.~Bokov \textit{et al.}, Vertex and energy reconstruction in JUNO with machine learning methods. Nucl. Instrum. Meth. A \textbf{1010}, 165527 (2021). \href{https://doi.org/10.1016/j.nima.2021.165527}{https://doi.org/10.1016/j.nima.2021.165527}

\bibitem{Li-2022nst}
Y.Z.~Li, Z.~Qian, J.H.~He \textit{et al.}, Improvement of machine learning-based vertex reconstruction for large liquid scintillator detectors with multiple types of PMTs. NUCL SCI TECH 33, 93 (2022). \href{https://doi.org/10.1007/s41365-022-01078-y}{https://doi.org/10.1007/s41365-022-01078-y}

\bibitem{Liu-2022nst}
H.L.~Liu, H.B.~Ji, J.M.~Zhang \textit{et al.}, A novel approach for feature extraction from a gamma-ray energy spectrum based on image descriptor transferring for radionuclide identification. NUCL SCI TECH 33, 158 (2022). \href{https://doi.org/10.1007/s41365-022-01150-7}{https://doi.org/10.1007/s41365-022-01150-7}

\end{thebibliography}
\end{document}